# Patient-level validation of a foundation-model pipeline for pediatric lung sounds: physician-annotated adventitious events are recognized, disease groups are not reliably predicted

Izzet Turkalp Akbasli[1*]
Oguzhan Serin[2]

[1] Department of Pediatric Intensive Care Medicine, Life Support Center, Hacettepe University, Ankara, Türkiye

[2] Department of Pediatric Emergency Medicine, Life Support Center, Hacettepe University Faculty of Medicine, Ankara, Türkiye

* Corresponding author: Izzet Turkalp Akbasli, MD, Department of Pediatric Intensive Care Medicine, Life Support Center, Hacettepe University, 06100 Sihhiye, Ankara, Türkiye. E-mail: iakbasli@hacettepe.edu.tr. ORCID: 0000-0003-3055-7355

Oguzhan Serin: oguzhanserin@hacettepe.edu.tr. ORCID: 0000-0002-3276-748X

## Abstract

**Background:** Lung-sound classifiers are usually evaluated on recording- or event-level splits, although each child contributes many recordings. We evaluated a foundation-model pipeline (PulmoVec) with the patient as the unit of partitioning and asked whether disease group can be predicted beyond age and sex.

**Methods:** We analyzed 19693 physician-annotated respiratory events from 736 children in the public SPRSound database. A frozen Health Acoustic Representations encoder with low-rank adapters was trained for screening (normal versus adventitious), sound pattern (normal, crackles, wheeze/rhonchi) and disease group (pneumonia, bronchial disease, normal/other); event probabilities were stacked with age, sex and auscultation site. The primary analysis was nested patient-grouped cross-validation in which every child was tested once, with a locked hold-out as secondary; comparators were the majority class, annotated event duration and demographics, and intervals came from bootstrapping patients.

**Results:** For physician-segmented events the area under the receiver operating characteristic curve (AUC) was 0.95 (95% CI 0.94 to 0.96) for screening and 0.95 (95% CI 0.94 to 0.96) for sound pattern, against 0.78 and 0.75 for event duration alone; the positive predictive value for adventitious events was 0.72 (95% CI 0.66 to 0.77). Disease-group prediction reached an AUC of 0.58 (95% CI 0.54 to 0.62); its accuracy (0.44) was below the majority-class rate (0.62) at event level and at patient level (0.46 versus 0.60), and age and sex carried 55.5% of the second-stage attribution for this outcome.

**Conclusions:** Recognition of physician-segmented adventitious events holds under patient-level validation and its discrimination is not explained by event duration or demographics, although in exploratory analyses its decisions favored short events and its positive predictive value would limit unaided use. Prediction of disease group, a clinical diagnosis, from lung-sound events alone was not demonstrated. Lung-sound studies should partition data by patient, as prior evidence indicates, and report non-acoustic comparators.

## Background

Chest auscultation is the first instrument used to assess a breathless child, yet lung sounds are described inconsistently [1, 2] and examiners agree only moderately on crackles and poorly on rhonchi [3]. Machine learning applied to electronic stethoscopes has been proposed as a way to make auscultation reproducible [4], and a meta-analysis of pediatric models pooled sensitivities and specificities near ninety percent [5], although reviews of the field stress its heterogeneity and the scarcity of external validation [6].

Such estimates depend on how data were partitioned; that meta-analysis rated overlap of children between training and test data as a risk of bias but did not stratify its pooled estimates by it. The SJTU Paediatric Respiratory Sound database (SPRSound), the pediatric resource used most often in that meta-analysis [5, 7], holds many recordings and annotated events per child, and part of its official test material comes from children who also appear in its training material [7, 8]. When one child's events fall on both sides of a split, a model can succeed by recognizing the child rather than the disease (identity confounding [9]), and record-wise validation overestimates clinical prediction performance [10, 11], including for lung sounds [12–14]. A further threat receives less attention: classifiers may exploit shortcuts [15], such as the annotated duration of the event for event-level classification [16] and age and sex for diagnosis [17].

PulmoVec is a multi-task stacking pipeline built on the Health Acoustic Representations (HeAR) foundation model [18]. Adventitious sounds are acoustic events that an examiner hears, whereas pneumonia and asthma are clinical diagnoses; whether the second can be read from the first is the question such a pipeline raises. Here we (i) estimate acoustic event classification by PulmoVec under patient-level validation against duration-only and demographics-only comparators, and (ii) test whether disease group can be predicted beyond demographics.

## Methods

### *Design and data*

The analysis plan, including the wording rules attached to the results, was written before the reported analyses were run and is kept, with a dated record of deviations, in the code repository; full methods are given in Additional file 1 (Section S1). This retrospective model development and internal validation study is reported according to TRIPOD+AI [19] (Additional file 1, Section S4). SPRSound contains recordings made with one electronic stethoscope at Shanghai Children's Medical Center, with event onsets, offsets and types annotated independently by three pediatric physicians (disagreements adjudicated by two senior physicians; no agreement statistic reported) and a recording-level diagnosis [7]. We used all annotated events with audio, excluding events without a usable label (stridor and "no event” annotations), recordings without a documented diagnosis and events with an implausible encoded age. Nothing was imputed. The base-model architecture, its training configuration and the disease grouping had been chosen during earlier development on SPRSound, in which the locked hold-out had been inspected; the analysis plan, dated 19 September 2026 and kept in the repository, therefore fixes the evaluation and its wording rules, not these pipeline choices. Patients and the public were not involved in this secondary analysis.

***Outcomes***

Three outcomes were modeled. *Screening* distinguished normal events from events with any adventitious sound; *sound pattern* distinguished normal events, crackles (including events with both wheeze and crackles) and wheeze/rhonchi; and *disease group* distinguished pneumonia, bronchial disease (bronchitis, asthma and bronchiolitis, grouped because their auscultatory signs overlap) and normal/other; the grouping was fixed before the analysis, and the recorded diagnoses in each group are listed in Additional file 1 (Section S2). The first two describe the event; the disease group belongs to the recording and is inherited by each of its events.

***Model***

Each event was cut at its annotated boundaries, centered in a 2 s clip whose margins were padded with low-level band-limited noise, band-pass filtered (100–1800 Hz) and encoded by the frozen HeAR Vision Transformer [18, 20] with low-rank adapters in its last six blocks [21]. One classifier per outcome was trained with focal loss [22], and their probabilities were combined with age, sex and auscultation site in a class-balanced LightGBM second stage [23, 24] trained only on out-of-fold probabilities and tuned with Optuna [25]. Full detail is in Additional file 1 (Sections S1 and S3).

***Evaluation***

The patient was always the unit of partitioning, and a program assertion verified before every fit and evaluation that no patient was shared between partitions. The primary analysis was nested patient-grouped cross-validation: five outer folds, so that each child was tested once, each with a validation partition and four-fold out-of-fold stacking inside the training patients [26, 27]. A locked hold-out, stratified by disease group, was secondary because it had been inspected once during development; it served model-specific analyses.

***Comparators and statistics***

We compared the full stack with the majority class and with second-stage models restricted to annotated event duration, to demographics, to the task’s own base model, and to all acoustic probabilities. Because a child’s events are correlated, all intervals are percentile intervals from 2000 bootstrap resamples of patients [28]; pre-specified paired contrasts used the same resampled patients, with Holm-adjusted $p$ values (the smallest attainable value, 0.0025 after adjustment, is reported as $p < 0.003$). We report the area under the receiver operating characteristic curve (AUC; macro-averaged for three classes), accuracy, per-class measures and calibration [29], at event level and after confidence-weighted aggregation to recordings and patients. Two wording rules were fixed in advance: acoustic classification is called robust only if the lower confidence bound of its AUC exceeds the duration-only AUC; disease-group prediction is reported as not demonstrated unless the full stack exceeds demographics (adjusted $p < 0.05$) and patient-level accuracy exceeds the majority-class rate.

### *Attribution and preprocessing controls*

After the planned analyses, and not pre-specified, two cues belonging to the preprocessing rather than to the sound were tested by retraining one base model per acoustic outcome on the locked split: the first with the mel bands above our filter's cut-off replaced by the clip median, the second with the event not isolated but taken as a 2 s window of real surrounding audio, so that the padded fraction no longer encodes the annotated duration. Child-level count rules, a second stage without class weighting and a child-level decision curve were also examined post hoc. Second-stage models were explained with Shapley additive explanations (SHAP) [30]; base models with patch-occlusion maps on the spectrogram input [31] for events sampled by a fixed rule, checked against deletion curves [32], a classifier-randomization test [33] and integrated gradients [34].

### *Use of generative artificial intelligence*

A generative artificial intelligence coding assistant (Claude Code, Anthropic) was used to help write and edit the analysis code, and generative artificial intelligence was also used for language editing. The analyses were specified by the authors, the code and its outputs were reviewed by the authors, and all analytical decisions were made by the authors; the study design and the interpretation of the results are the authors' own. The code enforces patient-level separation with assertions, and reported results were produced from saved predictions by scripts in the repository. The assistant does not meet the journal's authorship criteria and is not listed as an author. The authors take full responsibility for the code, the analyses and the content of the manuscript.

## Results

### *Cohort*

Of 24808 annotated events, 304 had no usable label, 4802 lacked a documented diagnosis (excluded from all outcomes so that the three share one cohort; 1377 of them also lacked a participant identifier) and 9 had an implausible age, leaving 19693 events in 4985 recordings from 736 children (Fig. 1, Table 1). A child contributed a median of 18 events, so event-level partitioning places almost every child on both sides of a split.

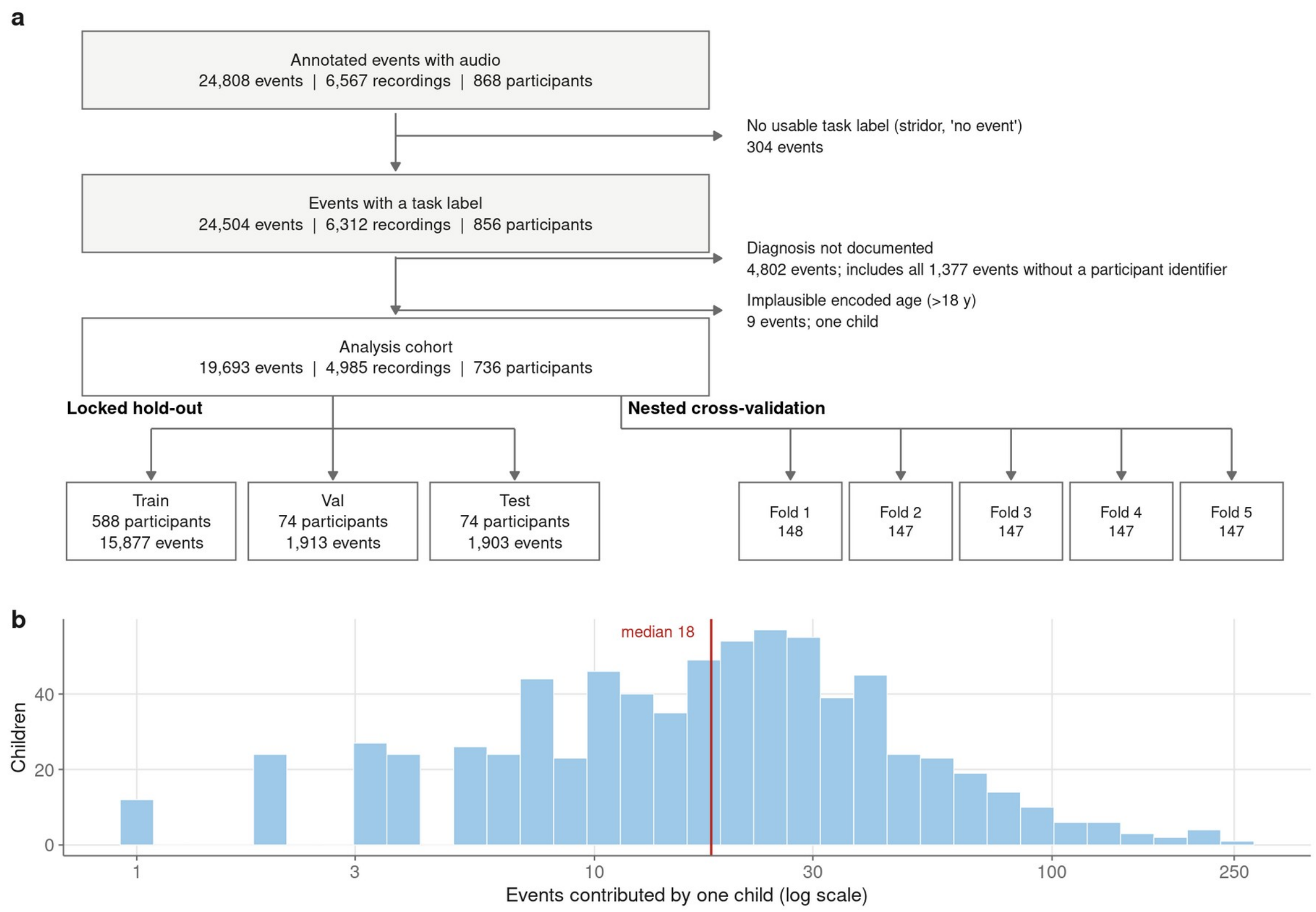

**Fig. 1 Flow of events, recordings and participants, and the number of events per child. a** Exclusions are applied to events; counts of recordings and participants are given at each stage. The analysis cohort is then partitioned in two ways: a locked hold-out, used as the secondary analysis, and a five-fold nested patient-grouped cross-validation, used as the primary analysis, in which each participant is tested exactly once. Within an outer fold the remaining participants form a validation partition and a training partition, the latter carrying four-fold patient-grouped out-of-fold stacking. **b** Distribution of the number of annotated events contributed by each child (logarithmic scale; red line, median). Because one child contributes many events, a split drawn on events leaves almost every child on both sides of it, whereas a split drawn on children cannot.

**Table 1 Cohort characteristics and the partitions of the locked hold-out**

| Variable | All | Train | Validation | Test | p (train vs test) |
|---|---|---|---|---|---|
| **Patients** | 736 | 588 | 74 | 74 | |
| Age (years), median (IQR) | 4.9 (3.4–7.2) | 4.8 (3.4–7.2) | 5.5 (3.3–8.3) | 5.0 (3.5–7.0) | 0.702 |
| Sex, n (%) | | | | | 0.519 |
| Female | 362 | 290 (49.3) | 32 (43.2) | 40 (54.1) | |
| Male | 374 | 298 (50.7) | 42 (56.8) | 34 (45.9) | |
| Disease group (patient-level), n (%) | | | | | 1.000 |
| Pneumonia | 438 | 350 (59.5) | 44 (59.5) | 44 (59.5) | |
| Bronchial diseases | 179 | 143 (24.3) | 18 (24.3) | 18 (24.3) | |
| Normal / Other | 119 | 95 (16.2) | 12 (16.2) | 12 (16.2) | |
| **Events** | 19693 | 15877 | 1913 | 1903 | |
| Event type, n (%) | | | | | <0.001 |
| Normal | 16396 | 13200 (83.1) | 1557 (81.4) | 1639 (86.1) | |
| Fine Crackle | 1692 | 1382 (8.7) | 196 (10.2) | 114 (6.0) | |
| Wheeze | 1243 | 1001 (6.3) | 131 (6.8) | 111 (5.8) | |
| Rhonchi | 208 | 172 (1.1) | 18 (0.9) | 18 (0.9) | |
| Coarse Crackle | 112 | 89 (0.6) | 10 (0.5) | 13 (0.7) | |
| Wheeze+Crackle | 42 | 33 (0.2) | 1 (0.1) | 8 (0.4) | |
| Auscultation site, n (%) | | | | | 0.607 |
| Left posterior (p1) | 4981 | 4011 (25.3) | 492 (25.7) | 478 (25.1) | |
| Left lateral (p2) | 5329 | 4335 (27.3) | 497 (26.0) | 497 (26.1) | |
| Right posterior (p3) | 4562 | 3669 (23.1) | 432 (22.6) | 461 (24.2) | |
| Right lateral (p4) | 4821 | 3862 (24.3) | 492 (25.7) | 467 (24.5) | |

Values are counts (%) unless stated. Disease group is a property of the patient; event type and auscultation site are counted per event. P values compare the training and test partitions (Mann–Whitney U test for age, chi-square test otherwise); the comparison of event types involves sparse cells.

### *Acoustic event classification*

Under nested cross-validation the AUC was 0.948 for screening and 0.950 for sound pattern (Table 2, Fig. 2), with between-fold ranges of 0.908 to 0.961 and 0.917 to 0.967; the 74-child hold-out gave 0.965 and 0.963. Adventitious events were detected with sensitivity 0.80 and specificity 0.94, but, with 3297 adventitious events among 19693 (16.7%), the positive predictive value was 0.72 (95% CI 0.66 to 0.77); crackles were hardest (sensitivity 0.70; Additional file 1, Table S1). AUC varied little by age band, sex or site (0.928 to 0.959 for screening; Additional file 1, Section S5), but sensitivity did: from 0.89 in children younger than 3 years to 0.65 in those aged 6 to 11 years, and 0.89 to 0.53 from the shortest to the longest tertile of event duration. Discrimination remained at or above 0.90 within tertiles, but the operating point shifted: specificity rose from 0.87 to 0.98 as sensitivity fell, 71% of adventitious events lay in the shortest tertile, and a threshold matched within each tertile to the overall specificity recovered only part of the loss for long events (0.68), while long crackles were detected at 0.27 (Additional file 1, Section S2). The class-balanced second stage overstated the probability of an adventitious sound (calibration slope 0.71, intercept −1.00).

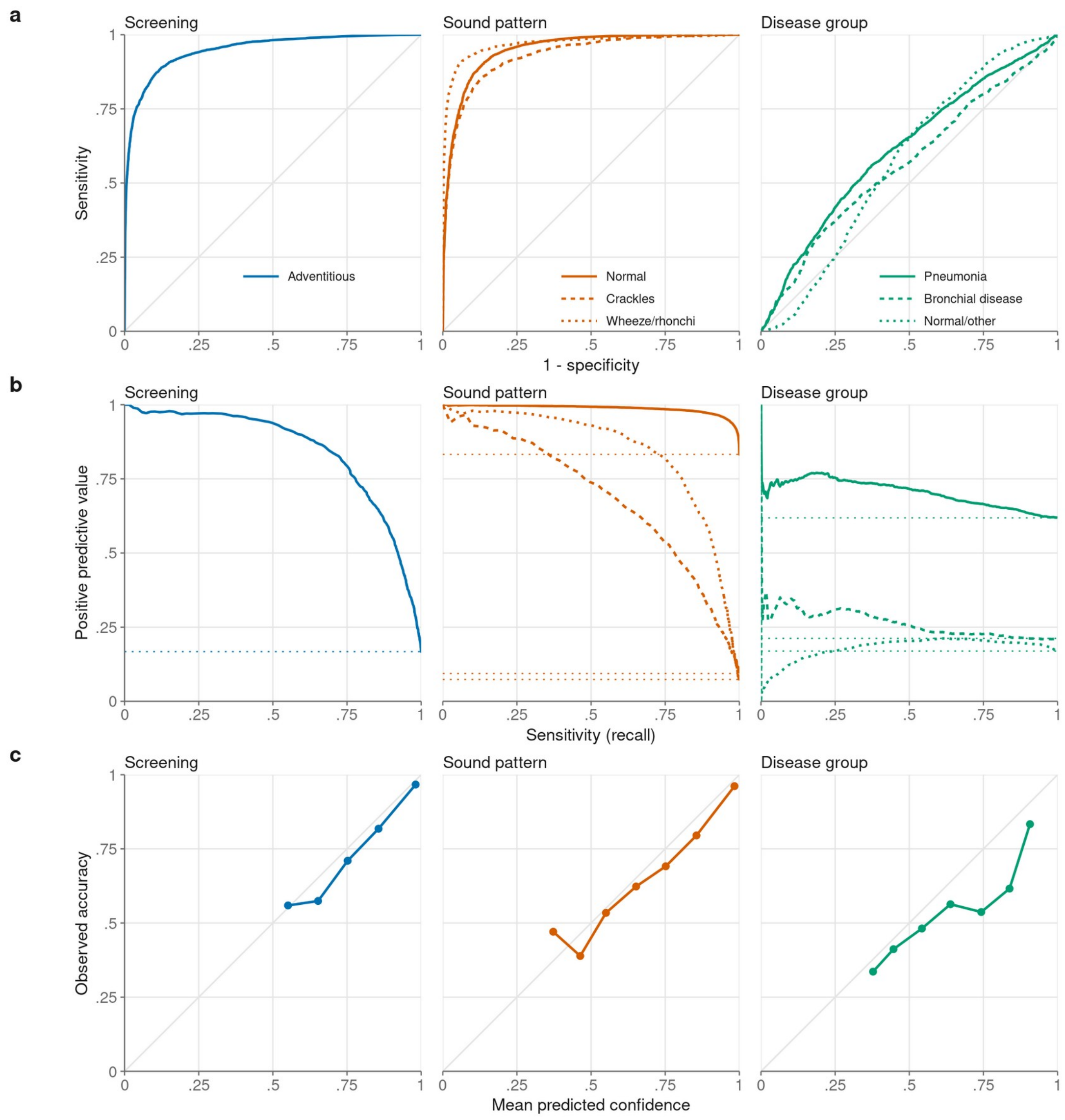

**Fig. 2 Discrimination and calibration under patient-level validation.** Pooled out-of-fold predictions of the nested cross-validation for the full stacked model. Color identifies the outcome; line type identifies the class within an outcome, keyed in the panels of **a**. **a** Receiver operating characteristic curves; the corresponding areas under the curve with intervals are given in Table 2. **b** Precision–recall curves; the dotted horizontal lines mark class prevalence, which is the performance of an uninformative model. **c** Reliability curves: predicted confidence against observed accuracy, in ten bins. Points below the diagonal indicate over-confidence, which is what the class-balanced second stage produces for screening.

At the child level the pre-specified confidence-weighted average missed most children with an adventitious sound (sensitivity 0.28, specificity 0.99), because a child's events are mostly normal while the derived label is "any adventitious event"; calling a child positive when at least two events were predicted adventitious, a post hoc rule, gave a sensitivity of 0.79 and a specificity of 0.78 (Additional file 1, Table S3); at a prevalence of 10%, the positive predictive value would be 0.29 for children under this rule and 0.58 for single events, and in a post hoc decision-curve analysis the count rule gave a higher net benefit than treating all children from a threshold probability of about 0.11 upward and exceeded the pre-specified rule at every threshold from 0.05 to 0.40 (Additional file 1, Section S2).

**Table 2 Performance of the full stacked model under patient-level validation**

| Outcome and level | Units | Majority-class accuracy | Accuracy | MCC | AUC, nested CV | AUC, hold-out |
|---|---|---|---|---|---|---|
| Screening, event | 19693 | 0.83 | 0.91 (0.91–0.92) | 0.71 (0.66–0.75) | 0.95 (0.94–0.96) | 0.97 (0.94–0.98) |
| Screening, patient[a] | 736 | 0.67 | 0.76 (0.73–0.79) | 0.44 (0.38–0.50) | 0.87 (0.84–0.90) | 0.89 (0.81–0.96) |
| Sound pattern, event | 19693 | 0.83 | 0.90 (0.89–0.91) | 0.69 (0.65–0.73) | 0.95 (0.94–0.96) | 0.96 (0.94–0.97) |
| Sound pattern, patient[a] | 736 | 0.67 | 0.74 (0.71–0.77) | 0.40 (0.34–0.46) | 0.88 (0.86–0.90) | 0.84 (0.76–0.91) |
| Disease group, event | 19693 | 0.62 | 0.44 (0.40–0.48) | 0.10 (0.05–0.16) | 0.58 (0.54–0.62) | 0.70 (0.57–0.80) |
| Disease group, patient | 736 | 0.60 | 0.46 (0.42–0.49) | 0.14 (0.09–0.19) | 0.60 (0.57–0.64) | 0.62 (0.50–0.74) |

Nested CV (cross-validation): pooled out-of-fold predictions of the five outer folds. Intervals are percentile bootstrap intervals obtained by resampling patients. AUC is macro-averaged one-vs-rest for three-class outcomes. MCC: Matthews correlation coefficient. Macro-F1 and further metrics are given in Additional file 1, Section S6. [a] Patient-level labels of the two acoustic outcomes are derived (any adventitious event; most frequent adventitious pattern), not annotated.

### *Comparators and shortcuts*

Annotated event duration alone reached an AUC of 0.779 for screening and 0.748 for sound pattern (Table 3, Fig. 3); the acoustic models exceeded it by 0.172 (95% CI 0.144 to 0.206) and 0.200 (95% CI 0.173 to 0.231) (adjusted $p < 0.003$), meeting the pre-specified condition for robustness, and kept AUCs of 0.902 or more within duration tertiles. Demographics alone were weakly informative (0.653 and 0.619). Stacking improved accuracy for both outcomes (by 0.018 and 0.055 against the task's own base model) and AUC for sound pattern (by 0.014), but lowered screening AUC (by 0.012); nine second-stage learners were near-identical (Additional file 1, Section S7).

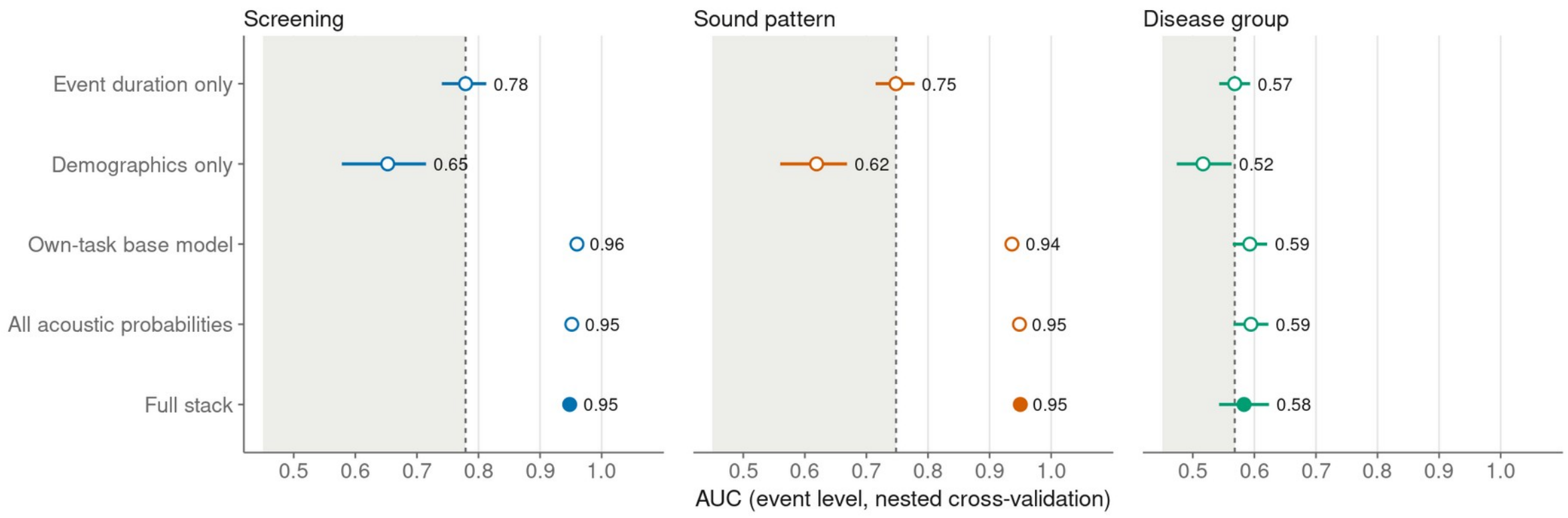

**Fig. 3 What the acoustic models add over two shortcuts.** Event-level area under the curve of second-stage models restricted to successive feature sets, from the nested cross-validation; bars are patient-bootstrap intervals, filled circles mark the full stack. All rows use the same learner and tuning protocol. The shaded region and the dashed line mark the performance reached by the annotated event duration alone, a feature that is a property of the annotation

rather than of the sound. The two acoustic outcomes lie far to the right of that line; for disease group the whole ladder lies within or at the edge of it, and the lower bound of the full stack does not clear it.

**Table 3 Comparator ladder: AUC at event level and paired contrasts (nested cross-validation)**

| Model | Screening | Sound pattern | Disease group |
|---|---|---|---|
| Event duration only | 0.779 (0.740–0.813) | 0.748 (0.715–0.778) | 0.568 (0.543–0.593) |
| Demographics only | 0.653 (0.578–0.715) | 0.619 (0.560–0.669) | 0.517 (0.474–0.563) |
| Own-task base model | 0.960 (0.951–0.968) | 0.936 (0.925–0.947) | 0.593 (0.565–0.621) |
| All acoustic probabilities | 0.952 (0.941–0.961) | 0.949 (0.940–0.957) | 0.594 (0.566–0.623) |
| Full stack | 0.948 (0.936–0.958) | 0.950 (0.940–0.958) | 0.583 (0.543–0.624) |
| *Δ acoustic − duration* | 0.172; $p < 0.003$ | 0.200; $p < 0.003$ | 0.026; $p = 0.255$ |
| *Δ full − demographics* | 0.295; $p < 0.003$ | 0.331; $p < 0.003$ | 0.067; $p < 0.003$ |
| *Δ full − own-task base* | −0.012; $p < 0.003$ | 0.014; $p < 0.003$ | −0.009; $p = 1.000$ |

All rows use the same second-stage learner and tuning protocol. Contrasts are paired patient-bootstrap differences with Holm-adjusted two-sided *p* values.

### *Disease-group prediction*

The full stack reached an AUC of 0.583 (Table 3; outer folds 0.477 to 0.677, Additional file 1, Table S13), and its accuracy of 0.44 was below the 0.62 obtained by always predicting pneumonia; at patient level accuracy was 0.46 against 0.60 (Table 2). The signal was not zero: the full stack exceeded demographics, the contrast used by the decision rule, by 0.067 (Table 3), and the acoustic probabilities alone exceeded them by 0.078 (95% CI 0.027 to 0.127) (adjusted $p = 0.020$), and patient-level balanced accuracy was 0.41 (95% CI 0.37 to 0.45) against a chance value of one third. It was, however, far too small to be useful and did not exceed annotated event duration (adjusted $p = 0.255$). The first pre-specified condition was met and the second was not, so disease-group prediction is reported as not demonstrated. Because the class-balanced second stage trades accuracy for minority-class recall, the accuracy rule is demanding, but balanced accuracy (0.39 at event level) and the Matthews correlation coefficient (0.10 and 0.14 at event and patient level) were also low; without class weighting, in a post hoc sensitivity analysis, accuracy rose to 0.60 at event level and 0.58 at patient level, still below the majority-class rates, and no pre-specified wording rule changed. Across 42 post hoc configurations at patient and recording level (36 models fitted without class weighting and 6 unfitted mean predicted probabilities), the patient-level AUC of the fitted models reached at most 0.65 for the three groups, with accuracy within 0.01 of the majority-class rate, 0.65 for pneumonia versus other diagnoses, with accuracy of up to 0.65 against 0.60, and 0.63 for pneumonia versus healthy controls (0.645 for the unfitted mean predicted probability of pneumonia), with accuracy at the majority-class rate (recording-level AUC up to 0.66, 0.69 and 0.70, respectively; Additional file 1, Section S2). Age and sex carried 55.5% of the second-stage attribution, against 2.8% for screening (Additional file 1, Section S8).

### *Attribution*

Removing the most attributed patches lowered the probability of the true class faster than random removal (difference in mean area under the deletion curve 0.114 for screening and 0.182 for sound pattern), but individual screening maps agreed no better with integrated gradients than with maps from a randomized classification head, so they should be read with caution (Additional file 1, Section S9). For adventitious events, 40.7% of attribution lay in one mel band of about 280 to 620 Hz (Fig. 4); for normal events in the screening model, 67.9% lay above our filter's cut-off, where only attenuated residual energy remains.

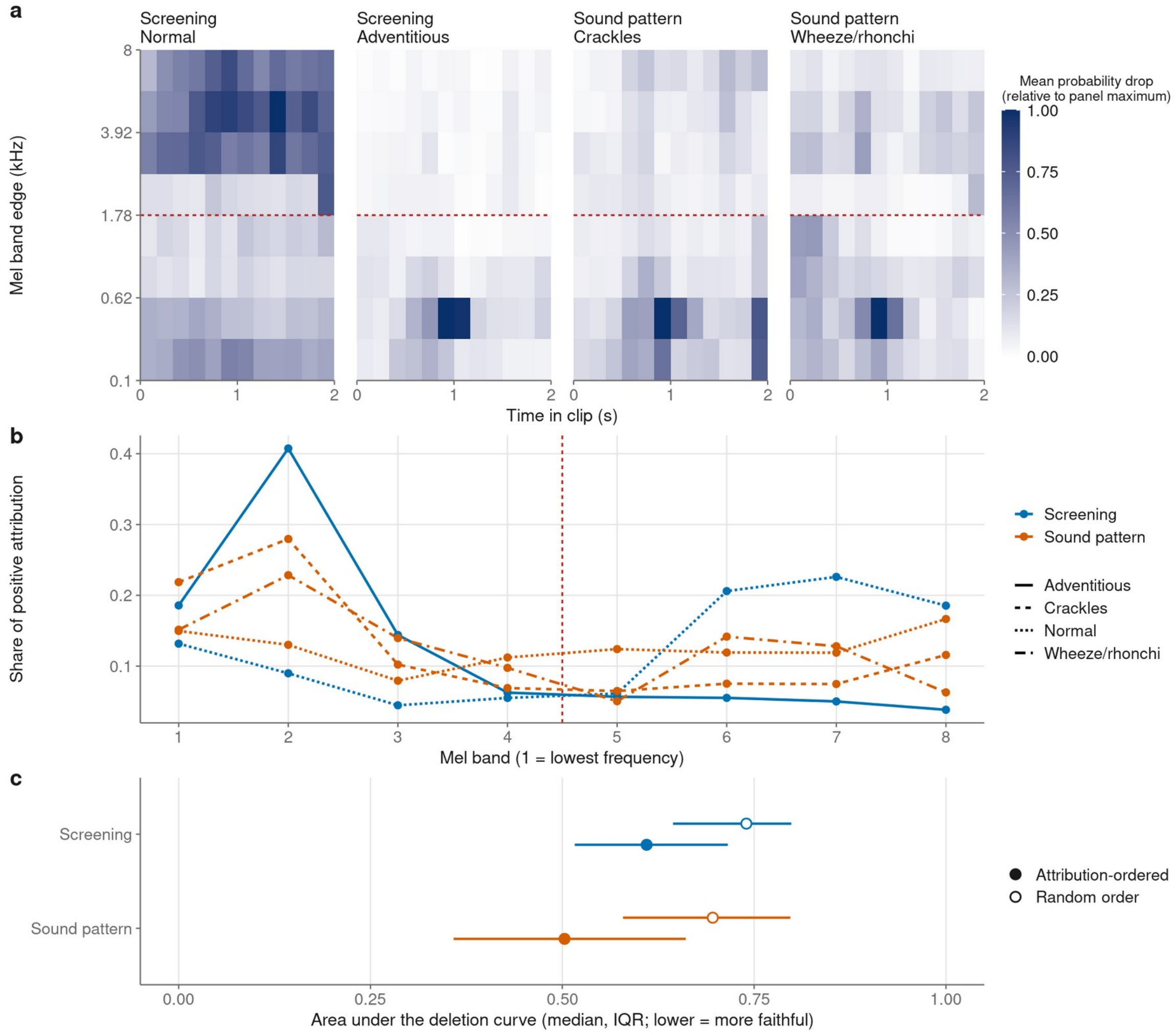

**Fig. 4 Model-derived attribution for the two acoustic outcomes. a** Class-mean occlusion maps over the twelve-by-eight patch grid of the encoder input (time by mel band, with band edges in kilohertz) for correctly classified hold-out events, scaled to the maximum of each panel; the dashed line marks the upper cut-off of the band-pass filter applied during preprocessing, above which only attenuated residual energy remains. **b** Share of positive attribution by mel band; color identifies the outcome and line type the class. Attribution for adventitious events concentrates in a low band, whereas attribution for normal events in the screening model lies largely above the filter cut-off; retraining without that region did not change discrimination (see Results). **c** Area under the deletion curve, median and interquartile range, when patches are removed in order of attribution and in random order; a lower value for attribution-ordered removal indicates that the maps identify regions the model actually uses.

Removing either preprocessing cue did not measurably change discrimination. Against hold-out base-model AUCs of 0.967 (screening) and 0.959 (sound pattern), masking the bands

above the cut-off changed the AUC by 0.002 (95% CI −0.001 to 0.007) and −0.001 (95% CI −0.008 to 0.010), and replacing the isolated clip with real surrounding audio by −0.006 (95% CI −0.013 to 0.003) and −0.006 (95% CI −0.016 to 0.014) (all $p > 0.1$); context clips admit neighboring events, so the second control may introduce another cue.

## Discussion

With the patient as the unit of partitioning, recognition of physician-segmented adventitious events held: an AUC of about 0.95, far above what annotated event duration achieves, independent of demographics, and barely changed on the hold-out when either preprocessing cue was removed. Sensitivity, however, varied with age and fell for long events: discrimination stayed at or above 0.90 within duration tertiles, but the operating point shifted toward short events. Disease group was not predictable to a clinically useful degree: the pooled interval excluded chance, though one outer fold fell below 0.5, and accuracy stayed below the majority-class rate.

That record-wise evaluation inflates performance is known [10, 11], including for lung sounds [12–14]. Our outcomes differ in their exposure to that mechanism: a diagnosis is constant within a child, so a model evaluated on a record-wise split can succeed by recognizing the child [9], whereas event type varies within a child and must be heard. Disease-level labels are therefore the most exposed to this bias, and the only other application of this foundation model to SPRSound that we found does not state its unit of partitioning [35]. On SPRSound itself, the best challenge models scored 0.10 to 0.15 lower on a later blind test set, a drop that may reflect new annotators and recording conditions as well as new children (Additional file 1, Section S10).

Two explanations, which this design cannot separate, fit the disease-group result. The first is clinical. Crackles and wheeze are properties of the sound itself, and a model trained on expert annotations learned to recognize them; pneumonia is a clinical diagnosis that integrates history, fever, respiratory rate, oxygen saturation and often imaging, no single examination finding reliably establishes it in children, and against radiographic pneumonia auscultatory findings weigh less than hypoxemia and increased work of breathing [36, 37]. In hospitalized children in low- and middle-income countries, digitally recorded crackles were not significantly associated with radiographic pneumonia [38]. The second is task construction: each event inherits its recording's diagnosis, so most acoustically normal events carry a disease label; the residual class is heterogeneous (71 of its 119 children were healthy controls); 27 children carried conflicting diagnoses between the source summary files; and the class-balanced second stage trades accuracy for minority-class recall, although removing the weighting did not change the result. DeepBreath likewise found pneumonia much harder than distinguishing healthy from pathological breathing [39], and a COVID-19 audio model was overtaken by a symptom questionnaire [17]. The post hoc child-level models, which assign each child a single label, also fell short of clinical usefulness, including for pneumonia versus healthy controls; this is consistent with the first explanation but does not establish it, because their inputs summarize three classifiers' outputs rather than the full acoustic representation. The finding concerns this labeling and these inputs, not lung sounds in general. Nor does it close the question: the acoustic probabilities carried a small signal beyond age and sex, and whether a child's aggregated lung sounds add to the clinical findings physicians already use deserves a prospective multicenter study that records oxygen saturation, respiratory rate, fever and imaging alongside the recordings.

These estimates must not be read as performance on unsegmented recordings: the models classify events whose boundaries physicians drew, a task no clinician performs, and a multi-country pediatric study without such boundaries reported balanced accuracies near seventy percent [40]. The models could serve as a second reader or teaching aid on annotated recordings, or as the classification stage behind an automatic event detector; it is that combined, child-level system that needs prospective evaluation. At the child level the pre-specified averaging rule missed most affected children; counting events did far better, but that rule was chosen post hoc and needs confirmation. Stacking, the pipeline's defining design feature, added little: it improved accuracy but slightly lowered screening AUC, so discrimination rests on the adapted foundation model.

Limitations include a single center and device, no external validation, and dependence on expert segmentation. Nor was a cross-release validation performed; it is feasible, but the releases differ sharply (adventitious events made up 24.1% of annotated events in the 2022 release and 5.3% in the 2024 release, with median annotated durations of 1.2 and 2.0 s), so its result would mix new children with differences in annotation and recording (Additional file 1, Section S10). The hold-out had been inspected before the analysis plan was written, so nested cross-validation is primary; its pooled bootstrap intervals are somewhat narrow, which is why fold ranges are given. Child-level acoustic labels are derived, the rarest event types were few, and probabilities need recalibration. The single-model ablations exclude large effects, not small ones. No sample size was calculated; the 736 children bound the disease-group AUC to within about ±0.04, enough to exclude an AUC above 0.64 for the pre-specified model; post hoc configurations reached 0.65 at patient level and 0.70 at recording level, with upper confidence bounds up to 0.71 and 0.75. The 1800 Hz cut-off may attenuate high-frequency components of crackles, the hardest class. The value of acoustic classifiers in primary care and low-resource settings is untested, and replication and deployment depend on access to the HeAR weights, which are not redistributable. Researchers should partition by patient and assert it in code, state the unit of partitioning in the abstract, and report demographics-only and duration-only comparators.

## Conclusions

A foundation-model pipeline recognizes physician-segmented adventitious lung-sound events in children with high discrimination under patient-level validation, with discrimination not explained by event duration or demographics, though its preference for short events, its positive predictive value and the child-level sensitivity of the pre-specified averaging rule, would limit its use unaided. Prediction of disease group from lung-sound events was not demonstrated, consistent with diagnoses that rest on clinical findings beyond auscultation; whether lung sounds add to those findings remains to be tested. Patient-level partitioning, which prior work shows to be necessary, and non-acoustic comparators, which here reached an AUC near 0.78, should be standard in lung-sound studies.

## List of abbreviations

AUC: area under the receiver operating characteristic curve; CI: confidence interval; CV: cross-validation; HeAR: Health Acoustic Representations; MCC: Matthews correlation coefficient; SHAP: Shapley additive explanations; SPRSound: SJTU Paediatric Respiratory Sound database; TRIPOD+AI: Transparent Reporting of a multivariable prediction model for Individual Prognosis Or Diagnosis, artificial intelligence extension.

## Declarations

### *Ethics approval and consent to participate*

This study is a retrospective secondary analysis of the publicly available, fully de-identified SPRSound database, which was collected at Shanghai Children’s Medical Center with approval from its Institutional Review Board (approval No. SCMCIRB-K2019056-1) [7]. Because only publicly available, de-identified data were used, no additional ethics committee approval or informed consent was sought; the study was conducted in accordance with the Declaration of Helsinki.

### *Consent for publication*

Not applicable.

### *Availability of data and materials*

The dataset supporting the conclusions of this article is available in the SPRSound repository, https://github.com/SJTU-YONGFU-RESEARCH-GRP/SPRSound [7, 8]. Participant-level partition assignments are provided in the PulmoVec repository (results_clean/nested_cv/patient_fold_assignment.csv and results_clean/holdout_patient_assignment.csv). Project name: PulmoVec. Project home page: https://github.com/turkalpmd/PulmoVec. Archived version: commit 013a6e5 of the repository above. Operating system: Linux. Programming language: Python 3.12. Other requirements: PyTorch 2.9, LightGBM, Optuna, SHAP; the HeAR encoder must be obtained from Google under the Health AI Developer Foundations terms [41] and is not redistributed. License: MIT. Restrictions to use by non-academics: those of the HeAR terms of use.

### *Competing interests*

The authors declare that they have no competing interests.

### *Funding*

This research received no specific grant from any funding agency in the public, commercial or not-for-profit sectors.

### *Authors’ contributions*

ITA conceived the study, designed the methodology, developed the software, performed the analysis, curated the data, prepared the figures and led the writing of the manuscript. OS contributed to validation and supervision and revised the manuscript critically. Both authors read and approved the final manuscript.

### *Acknowledgements*

We thank the SPRSound investigators for making their data public. A preprint of this article is available on arXiv (arXiv:2603.15688).

**Additional files**

**Additional file 1** (.docx). **Supplementary methods, results, reporting checklist and data.** S1, supplementary methods; S2, supplementary results (per-class performance, aggregation, subgroups, calibration, attribution, preprocessing ablations, child-level aggregation rules, recorded diagnoses within disease groups, post hoc disease-group models, a second stage without class weighting and a child-level decision curve); S3, training configuration and search space; S4, TRIPOD+AI checklist; S5, subgroup performance by age band, sex, auscultation site and event-duration tertile; S6, further metrics of the primary analysis at event level, after aggregation to recordings and patients, and by outer fold; S7, benchmark of nine second-stage learners on the hold-out; S8, SHAP attribution of the second-stage learners; S9, further occlusion and integrated-gradients maps, selected by the fixed rule, and the classifier-randomization test; S10, composition of the cohort by SPRSound release and published SPRSound event-classification results.

# Additional file 1

## Supplementary methods, results, reporting checklist and data

*Patient-level validation of a foundation-model pipeline for pediatric lung sounds: physician-annotated adventitious events are recognized, disease groups are not reliably predicted*

This file brings together all supplementary material for the article in one document. Tables and figures numbered without the prefix S refer to the main article; tables that appear in the main article are not repeated here. Every number is generated from the result files of the public code repository. In Sections S5 to S7, values are rounded to three decimal places.

**Contents**

# S1 Supplementary methods

## Study design and reporting

This is a retrospective model development and internal validation study using a public, de-identified dataset. It is reported according to TRIPOD+AI [1] (Section S4). The analysis plan (primary and secondary evaluation designs, comparators, contrasts, subgroups and the wording rules attached to the results) was written before the analyses reported here were run and is kept, with a dated record of deviations, in the code repository. Analyses added afterwards are marked as post hoc wherever they are reported.

## Data source

SPRSound comprises lung-sound recordings collected with an electronic stethoscope at Shanghai Children's Medical Center, with event-level annotations (onset, offset and type) and a recording-level diagnosis [2, 3]. Each record of adequate signal quality was segmented and annotated independently by three pediatric physicians; disagreements were adjudicated by two senior physicians, and no inter-annotator agreement statistic is reported [2]. We assembled all annotated events from the classification and detection releases for which the audio file was available.

## Participants and eligibility

Events were excluded if they had no usable task label (stridor and "no event" annotations), if the diagnosis of the recording was not documented ("Unknown"), or if the age encoded for the recording was implausible for a pediatric cohort (above 18 years). The last rule removed nine events of one child whose other recordings carried an age of 2.5 years. Undocumented diagnoses were excluded from the whole study rather than folded into a residual disease class, so that all three tasks are evaluated on one cohort. The flow of events and participants is shown in Fig. 1. Recording-level diagnoses come from three participant summary tables of the database; where they disagreed (27 children in the cohort), the Grand Challenge 2024 table took precedence, then the TBioCAS table, then the Grand Challenge 2023 table. No data were imputed.

## Outcomes

Three outcomes were defined in advance. *Screening*: normal versus any adventitious sound. *Sound pattern*: normal, crackles (fine crackle, coarse crackle, wheeze with crackle) or wheeze/rhonchi. *Disease group*: pneumonia (severe or non-severe), bronchial disease (bronchitis, asthma, bronchiolitis) or normal/other. The first two are properties of the annotated event. The disease group is a property of the recording and is inherited by every event in it, including normal-sounding events of a child with pneumonia; this is a deliberate feature of the question being asked (can disease be read from events?) and a limitation we return to in the Discussion.

## Audio preprocessing

Recordings were resampled to 16 kHz mono. Each annotated event was cut at its annotated boundaries, tapered with a 10 ms raised-cosine window, and placed at the center of a 2 s buffer, the input length of the encoder. The margins were filled with low-level band-limited noise (100–600 Hz, scaled to 8% of the event's root-mean-square amplitude) rather than silence or neighboring audio, so that adjacent events cannot contaminate the clip. Events longer than 2 s were center-cropped. The clip was band-pass filtered (fourth-order Butterworth, 100–1800 Hz) and peak-normalized. During training only, the event position was jittered by up to ±200 ms.

Because the fraction of the clip occupied by the event reveals its annotated duration, duration was treated as a potential shortcut and analyzed explicitly (see Comparators).

## Model

The encoder is HeAR, a Vision Transformer (ViT-L) pre-trained by masked auto-encoding on health-related audio [4–6], used with its released preprocessing, which converts a 2 s clip to a 192 × 128 mel-spectrogram. We used the weights released on Hugging Face as google/hear-pytorch (revision f791cd42; unchanged since April 2025) and stored their SHA-256 digest with every run. All encoder weights were frozen. Low-rank adapters (LoRA; rank 8, scaling 16) [7] were attached to the query and value projections of the last six of 24 transformer blocks. The 512-dimensional pooled embedding was concatenated with a 1024-dimensional attention-pooled summary of the 96 patch tokens and passed to a three-layer classification head; 1.4% of the parameters were trainable. One such model was trained per outcome with focal loss ($\gamma = 2$, inverse-frequency class weights) [8], AdamW (learning rate $2 \times 10^{-4}$, weight decay 0.01), cosine schedule with two warm-up epochs, batch size 32, at most 25 epochs, and early stopping on validation macro-F1 (patience 6).

Event-level probabilities of the three base models (eight values) were combined with age, sex and auscultation site in a second-stage LightGBM classifier [9–11]. The second stage was trained only on out-of-fold base-model probabilities: within the training patients, base models were fitted on all but one patient-grouped fold and scored the held-out fold. The second stage used balanced class weights. Hyper-parameters were tuned with Optuna (tree-structured Parzen estimator, 100 trials) [12] to maximize macro-F1 on the validation patients; the search space is given in Table S8.

## Evaluation design

The unit of partitioning was always the patient. *Primary analysis*: nested patient-grouped cross-validation over the whole cohort. Patients were divided into five outer folds stratified by disease group, so that each child was a test patient exactly once. Within each outer fold, the remaining patients were split into a validation set (one tenth of the cohort) and a training set, on which four-fold patient-grouped out-of-fold stacking was performed; inner models were early-stopped on the validation set, never on the fold they subsequently scored [13, 14]. We report metrics on the pooled out-of-fold test predictions together with the mean, standard deviation and range across outer folds. *Secondary analysis*: a single locked hold-out (80/10/10% of patients). The hold-out had been inspected once during development and is therefore used for model-specific analyses (calibration plots, attribution), not as the primary estimate. A program assertion verified before every fit and every evaluation that no patient identifier was shared between partitions.

## Comparators

Performance was compared with (a) the majority-class rate; (b) a model using only the annotated event duration; (c) a model using only age, sex and auscultation site; (d) the task's own base model without stacking; (e) all acoustic probabilities without demographics; and (f) the full stack. Models (b)–(f) used the same second-stage learner and tuning protocol. To test whether the choice of second-stage learner matters, nine alternative learners were fitted to the same features (Section S7).

## Statistical analysis

Events from one child are correlated, so all uncertainty was estimated with a cluster bootstrap that resamples patients (2000 replicates, percentile 95% confidence intervals) [15]; DeLong's test was not used for the same reason. We report accuracy, balanced accuracy, macro-F1, Matthews correlation coefficient, area under the receiver operating characteristic curve (AUC; macro-averaged one-vs-rest for three-class tasks), area under the precision–recall curve, per-class sensitivity, specificity and predictive values, the Brier score, expected calibration error (ten bins) and, for the binary task, calibration slope and intercept [16, 17]. Pre-specified paired contrasts between comparators used the same resampled patients for both models; two-sided bootstrap *p* values were Holm-adjusted within task. Event-level probabilities were aggregated to recordings and to patients by confidence-weighted averaging (simple averaging as a sensitivity analysis). Recording- and patient-level labels for the two acoustic outcomes are derived ("any adventitious event"; most frequent adventitious pattern) and are described as such. Subgroup performance by age band, sex, auscultation site and event-duration tertile is exploratory; cells with fewer than ten patients were suppressed. Two wording rules were fixed in advance: acoustic classification would be called robust only if the lower confidence bound of its AUC exceeded the AUC of the duration-only model, and disease-group prediction would be reported as not demonstrated unless the full stack exceeded the demographics-only model (adjusted $p < 0.05$) and patient-level accuracy exceeded the majority-class rate.

## Model attribution

For the second stage we computed SHAP values with TreeExplainer on test events [18, 19] and summarized the share of mean absolute attribution carried by the task's own acoustic probabilities, the other tasks' probabilities, and each demographic variable. For the base models we computed patch-occlusion maps on the spectrogram input (each of the 96 patches replaced by the clip median; attribution = fall in the probability of the true class) [20], with integrated gradients [21] as a second method. Events were sampled by a fixed seeded rule (up to 150 per class), not selected by hand. Faithfulness was assessed with deletion curves against random patch order [22], a classifier-randomization test [23], the share of attribution falling inside the annotated event relative to the share of the clip the event occupies, and the distribution of attribution across mel bands. Attention-pooling weights are reported descriptively only, because one branch of the classifier bypasses the pooling layer.

## Reproducibility

Random seeds were fixed for Python, NumPy, PyTorch and the data-loader workers (seed 42); training used bfloat16 autocast on one NVIDIA RTX 3090, which leaves small non-deterministic differences between runs. The SHA-256 digest of the encoder weights, package versions and the code revision are stored with every result set. Every number in this article was generated programmatically from the result files.

# S2 Supplementary results

## Cohort

Of 24808 annotated events with audio, 304 had no usable task label, 4802 belonged to recordings without a documented diagnosis and 9 carried an implausible age, leaving 19693 events in 4985 recordings from 736 children (Fig. 1). All recordings without a participant identifier fell among those without a documented diagnosis, so every analyzed event could be assigned to a child. Children contributed a median of 18 events (interquartile range 9 to 33, maximum 254) and 5 recordings, which is why event-level partitioning places almost every child on both sides of a split. Table 1 describes the locked hold-out; age, sex, disease group and auscultation site were balanced between partitions, whereas the distribution of event types differed.

### Acoustic event classification under patient-level validation

In the nested cross-validation, the AUC was 0.948 (95% CI 0.936 to 0.958) for screening and 0.950 (95% CI 0.940 to 0.958) for sound pattern (Table 2, Fig. 2). Across the five outer folds the AUC ranged from 0.908 to 0.961 and from 0.917 to 0.967, respectively. The locked hold-out gave higher point estimates (0.965 and 0.963) that lay above the between-fold range for screening and within it for sound pattern, which illustrates how much a single test set of 74 children can flatter a model. Adventitious events were detected with a sensitivity of 0.80 (95% CI 0.76 to 0.85) and a specificity of 0.94 (95% CI 0.93 to 0.95); because only 3297 of 19693 events were adventitious, the positive predictive value was 0.72 (95% CI 0.66 to 0.77) (Table S1). Crackles were the hardest pattern (sensitivity 0.70 (95% CI 0.65 to 0.75), positive predictive value 0.59 (95% CI 0.51 to 0.66)).

**Table S1 Per-class performance of the full stacked model at event level (nested cross-validation)**

| Outcome | Class | Events | Sensitivity | Specificity | PPV | NPV |
|---|---|---|---|---|---|---|
| Screening | Adventitious | 3297 | 0.80 (0.76–0.85) | 0.94 (0.93–0.95) | 0.72 (0.66–0.77) | 0.96 (0.95–0.97) |
| Sound pattern | Normal | 16396 | 0.94 (0.93–0.94) | 0.81 (0.77–0.85) | 0.96 (0.95–0.97) | 0.72 (0.67–0.77) |
| Sound pattern | Crackles | 1846 | 0.70 (0.65–0.75) | 0.95 (0.94–0.96) | 0.59 (0.51–0.66) | 0.97 (0.96–0.98) |
| Sound pattern | Wheeze/rhonchi | 1451 | 0.80 (0.73–0.85) | 0.98 (0.97–0.98) | 0.75 (0.69–0.81) | 0.98 (0.98–0.99) |
| Disease group | Pneumonia | 12181 | 0.50 (0.44–0.55) | 0.67 (0.61–0.73) | 0.71 (0.64–0.77) | 0.45 (0.39–0.51) |
| Disease group | Bronchial disease | 4176 | 0.38 (0.30–0.46) | 0.73 (0.69–0.77) | 0.27 (0.21–0.34) | 0.81 (0.77–0.86) |
| Disease group | Normal/other | 3336 | 0.30 (0.22–0.39) | 0.73 (0.68–0.77) | 0.18 (0.12–0.26) | 0.84 (0.80–0.87) |

One-vs-rest; intervals by patient bootstrap. PPV and NPV: positive and negative predictive value.

### Comparators and shortcuts

Annotated event duration alone was informative: a model that saw nothing else reached an AUC of 0.779 for screening and 0.748 for sound pattern (Table 3, Fig. 3). The acoustic models exceeded it by 0.172 (95% CI 0.144 to 0.206) and 0.200 (95% CI 0.173 to 0.231) (adjusted $p <$ 0.003 for both), and within tertiles of event duration the AUC of the full model remained between 0.902 and 0.945 for screening and between 0.908 and 0.947 for sound pattern. The lower confidence bounds of the acoustic AUCs were far above the duration-only estimates, so the pre-specified condition for describing acoustic classification as robust to this shortcut was met. Demographics alone carried little information about event type (AUC 0.653 and 0.619), and adding them to the acoustic probabilities changed the AUC by −0.003 (95% CI −0.008 to 0.000) and 0.001 (95% CI −0.003 to 0.005).

Whether stacking helped depended on the outcome and the metric. Relative to the task's own base model, the full stack changed the AUC by −0.012 (95% CI −0.017 to −0.008) for screening (a loss; adjusted $p < 0.003$) and by 0.014 (95% CI 0.008 to 0.020) for sound pattern (a gain; adjusted $p < 0.003$), while accuracy improved for both (0.018 (95% CI 0.011 to 0.024) and 0.055 (95% CI 0.047 to 0.062)). The choice of second-stage learner was immaterial: on the hold-out, the AUC of nine second-stage learners ranged from 0.962 to 0.968 for screening and from 0.962 to 0.967 for sound pattern, with overlapping intervals (Section S7).

### Disease-group prediction

Disease group could not be predicted usefully from lung-sound events. The AUC of the full stack was 0.583 (95% CI 0.543 to 0.624), and its accuracy of 0.44 (95% CI 0.40 to 0.48) was below the 0.62 obtained by always predicting pneumonia. Aggregating a child's events did not rescue it: patient-level accuracy was 0.46 (95% CI 0.42 to 0.49) against a majority-class rate of 0.60, with an AUC of 0.602 (95% CI 0.569 to 0.636). The full stack exceeded the demographics-only model by 0.067 (95% CI 0.028 to 0.107) (adjusted $p < 0.003$), a difference that is statistically detectable in 736 children but too small to matter, and the acoustic probabilities did not exceed event duration alone (0.026 (95% CI −0.004 to 0.056); adjusted $p$ = 0.255). The first pre-specified condition (full stack better than demographics) was met and the second (patient-level accuracy above the majority-class rate) was not, so disease-group prediction is reported as not demonstrated. The acoustic probabilities alone also exceeded demographics (0.078 (95% CI 0.027 to 0.127); adjusted $p = 0.020$), and patient-level balanced accuracy was 0.41 (95% CI 0.37 to 0.45) against a chance value of one third. In the hold-out model, 55.5% of the mean absolute SHAP attribution of the disease-group learner fell on age and sex, compared with 2.8% and 9.5% for the two acoustic outcomes (Section S8). Between-fold variation was large (AUC 0.477 to 0.677), and the hold-out estimate of 0.704 (95% CI 0.571 to 0.797) shows how wide the uncertainty of a single small test set is.

### Aggregation, subgroups and calibration

Recording-level aggregation gave an AUC of 0.938 (95% CI 0.924 to 0.950) for the presence of any adventitious event in 4985 recordings, and patient-level aggregation 0.872 (95% CI 0.842 to 0.901); simple averaging gave similar results (Section S6). Event-level AUCs of the acoustic outcomes varied little by age band (0.928 to 0.959 for screening), sex (0.943 to 0.951) or auscultation site (0.934 to 0.957), but screening sensitivity ranged from 0.65 to 0.89 across age bands and from 0.53 to 0.89 across duration tertiles (Section S5). Expected calibration error was low (0.020 for screening, 0.027 for sound pattern), but the pooled screening probabilities were not well calibrated in the logistic sense (slope 0.71, intercept −1.00): the class-balanced second stage overstates the probability of an adventitious sound, and probabilities would need recalibration before any use as risks (Fig. 2).

### What the acoustic models attend to

Occlusion maps were computed for 300 and 414 hold-out events. Removing the most attributed patches first lowered the probability of the true class faster than removing patches at random (difference in area under the deletion curve 0.114 for screening and 0.182 for sound pattern; differences in mean area). Maps of the trained model correlated with maps from a model whose classification head had been re-initialized with median Spearman coefficients of 0.19 (screening) and 0.03 (sound pattern), and with integrated gradients of 0.09 and 0.13; for screening the agreement with the randomized head was the higher of the two, so individual maps should be read with caution. For events classified as adventitious, 40.7% of positive attribution lay in a single mel band spanning roughly 280 to 620 Hz, the range in which crackles and wheezes carry most of their energy (Fig. 4). For normal events, 67.9% of the screening model's attribution (52.9% for the sound-pattern model) lay in bands above about 1780 Hz, above the cut-off of our own band-pass filter, where only attenuated residual energy remains after per-channel gain normalization (20.0% for adventitious events). The models therefore read that region when it is available; the retraining ablation (Table S2) shows that they do not need it. The share of attribution inside the annotated event (0.67 for sound pattern, 0.80 for screening) was only modestly above the share of the clip that the event occupies (0.63 and 0.70). The maps are therefore presented as a faithfulness-checked description of spectral emphasis, not as localization of the pathological sound.

## Preprocessing ablations (post hoc)

One base model per acoustic outcome was retrained on the locked split with (i) the mel bands above the band-pass cut-off replaced by the clip median, and (ii) the event taken as a 2 s window of real surrounding audio instead of being isolated and noise-padded. The reference is the existing hold-out base model scored by the same code; the ablated models were trained afresh, so the differences include run-to-run variation. Context clips admit neighboring events, which may themselves carry label information, so the second ablation removes the duration cue but can introduce another.

**Table S2 Base-model AUC on the locked hold-out with each preprocessing cue removed**

| Outcome | Reference | Bands above cut-off masked | Real surrounding audio |
|---|---|---|---|
| Screening | 0.967 (0.943–0.981) | 0.969 (0.946–0.983) | 0.961 (0.936–0.978) |
| Sound pattern | 0.959 (0.936–0.971) | 0.958 (0.943–0.968) | 0.952 (0.940–0.966) |

Paired differences from the reference: screening 0.002 (95% CI −0.001 to 0.007) (masking) and −0.006 (95% CI −0.013 to 0.003) (context); sound pattern −0.001 (95% CI −0.008 to 0.010) and −0.006 (95% CI −0.016 to 0.014).

## Child-level aggregation rules (post hoc)

The pre-specified rule averages a child's event probabilities with confidence weights, whereas the derived child-level label is "any adventitious event"; because most of a child's events are normal, the average rarely favors an adventitious class. Two count rules that match the label were therefore compared post hoc.

**Table S3 Child-level detection of adventitious sounds by aggregation rule (nested cross-validation)**

| Outcome | Rule | Sensitivity, any adventitious | Specificity | Balanced accuracy |
|---|---|---|---|---|
| Screening | Confidence-weighted average (pre-specified) | 0.28 (0.23–0.34) | 0.99 (0.99–1.00) | 0.64 (0.61–0.67) |
| Screening | At least one event adventitious | 0.92 (0.88–0.95) | 0.57 (0.52–0.61) | 0.74 (0.72–0.77) |
| Screening | At least two events adventitious | 0.79 (0.74–0.84) | 0.78 (0.75–0.82) | 0.79 (0.76–0.82) |
| Sound pattern | Confidence-weighted average (pre-specified) | 0.26 (0.21–0.32) | 0.99 (0.99–1.00) | 0.48 (0.45–0.52) |
| Sound pattern | At least one event adventitious | 0.92 (0.88–0.95) | 0.56 (0.52–0.61) | 0.69 (0.65–0.73) |
| Sound pattern | At least two events adventitious | 0.81 (0.76–0.86) | 0.77 (0.73–0.80) | 0.69 (0.65–0.73) |

736 children, of whom 241 had at least one annotated adventitious event. Intervals by bootstrap over children.

## Recorded diagnoses and post hoc disease-group models (post hoc)

Table S4 lists the recorded diagnoses within each disease group. The pneumonia and bronchial groups each combine closely related diagnoses, whereas the normal/other group combines 71 healthy control children with 48 children carrying one of nine other diagnoses, from acute upper respiratory infection to bronchiectasis and pulmonary hemosiderosis; under the pre-specified grouping, one child with protracted bacterial bronchitis also falls in this group. Each diagnosis is a single recording-level label without supporting clinical data, so some label noise is to be expected.

**Table S4 Recorded diagnoses within each disease group**

| Disease group | Recorded diagnosis | Children | Recordings | Events |
|---|---|---|---|---|
| Pneumonia | Pneumonia (non-severe) | 398 | 2814 | 10919 |
| | Pneumonia (severe) | 40 | 304 | 1262 |
| Bronchial disease | Bronchitis | 105 | 564 | 2301 |
| | Asthma | 66 | 418 | 1570 |
| | Bronchiolitis | 8 | 48 | 305 |
| Normal/other | Healthy control | 71 | 456 | 1994 |
| | Other respiratory diseases | 15 | 96 | 381 |
| | Acute upper respiratory infection | 11 | 65 | 235 |
| | Bronchiectasis | 8 | 108 | 313 |
| | Chronic cough | 4 | 13 | 54 |
| | Hemoptysis | 4 | 61 | 219 |
| | Airway foreign body | 2 | 11 | 38 |
| | Pulmonary hemosiderosis | 2 | 20 | 70 |
| | Kawasaki disease | 1 | 3 | 11 |
| | Protracted bacterial bronchitis | 1 | 4 | 21 |

Analysis cohort (736 children, 4985 recordings, 19693 events). The disease group is assigned from the recorded diagnosis by a rule fixed before the analysis.

The pre-specified analysis classifies single events, although the diagnosis is shared by all events of a child. After the primary results were known, we therefore fitted models to each child's aggregated event probabilities. For every child, the probabilities of the three base models from the outer fold in which that child was tested were summarized (mean, extremes, standard deviation and quartiles of each probability, the fraction of events assigned to each predicted class, and the number and duration of events and the number of recordings), giving 61 acoustic features, to which age and sex were added. Logistic regression and a small LightGBM with hyper-parameters fixed in advance were evaluated with a new patient-grouped five-fold cross-validation repeated with five seeds, for the three disease groups, for pneumonia versus all other diagnoses and, as the clinically most coherent contrast, for pneumonia versus healthy controls (Table S5). None of these models used class weighting.

For the three groups, the patient-level AUC rose from 0.602 in the pre-specified analysis to 0.627 and 0.645, but accuracy (0.592 and 0.599) did not differ from the majority-class rate of 0.595, and balanced accuracy stayed below 0.39: the models predicted pneumonia for most children. For pneumonia versus other diagnoses, the AUC was 0.652 and 0.647 and accuracy 0.649 and 0.633, exceeding the majority-class rate by 0.055 (95% CI 0.016 to 0.094) and 0.039 (95% CI 0.000 to 0.079). The acoustic aggregates added 0.048 (95% CI 0.009 to 0.089) to the AUC of the LightGBM model with age and sex alone for the three groups and 0.034 (95% CI −0.015 to 0.080) for pneumonia versus other diagnoses. Recording-level models gave similar results (three groups: AUC up to 0.664, accuracy 0.634 against a majority-class rate of 0.625; pneumonia versus other diagnoses: AUC up to 0.687, accuracy up to 0.672; pneumonia versus healthy controls: AUC up to 0.698). Restricted to the 438 children with pneumonia and the 71 healthy controls, the patient-level AUC was 0.62 to 0.63 for the fitted models with acoustic features (0.645 for the unfitted mean predicted probability of pneumonia), and accuracy stayed at the majority-class rate of 0.861: the models called almost every child pneumonia. These analyses were designed after the primary results were known, and 42 configurations were evaluated without correction for multiplicity. They do not change the pre-specified conclusion: pneumonia can be separated from other diagnoses better than chance (AUC 0.65), but not to a degree that would support clinical use, and any such use would require external validation. The base-model probabilities that these models aggregate come from five outer-fold models whose calibration may differ, which adds variance but no information about the test children.

**Table S5 Post hoc patient-level disease-group models (nested cross-validation probabilities)**

| Target | Features | Learner | AUC (95% CI) | Accuracy (95% CI) |
|---|---|---|---:|---:|
| Three disease groups | Age and sex | Logistic regression | 0.537 (0.504–0.574) | 0.595 (0.560–0.629) |
| | Age and sex | LightGBM | 0.598 (0.565–0.632) | 0.594 (0.560–0.630) |
| | Acoustic aggregates | Logistic regression | 0.615 (0.581–0.650) | 0.601 (0.567–0.637) |
| | Acoustic aggregates | LightGBM | 0.630 (0.597–0.664) | 0.590 (0.554–0.624) |
| | Acoustic aggregates, age and sex | Logistic regression | 0.627 (0.593–0.663) | 0.592 (0.560–0.628) |
| | Acoustic aggregates, age and sex | LightGBM | 0.645 (0.612–0.680) | 0.599 (0.565–0.635) |
| Pneumonia vs other | Age and sex | Logistic regression | 0.517 (0.473–0.561) | 0.591 (0.556–0.624) |
| | Age and sex | LightGBM | 0.614 (0.571–0.654) | 0.620 (0.586–0.655) |
| | Acoustic aggregates | Logistic regression | 0.634 (0.592–0.674) | 0.607 (0.572–0.640) |
| | Acoustic aggregates | LightGBM | 0.628 (0.588–0.667) | 0.610 (0.577–0.644) |
| | Acoustic aggregates, age and sex | Logistic regression | 0.652 (0.611–0.693) | 0.649 (0.618–0.685) |
| | Acoustic aggregates, age and sex | LightGBM | 0.647 (0.607–0.687) | 0.633 (0.601–0.667) |
| Pneumonia vs healthy control | Age and sex | Logistic regression | 0.471 (0.406–0.543) | 0.861 (0.829–0.890) |
| | Age and sex | LightGBM | 0.521 (0.452–0.597) | 0.861 (0.829–0.890) |
| | Acoustic aggregates | Logistic regression | 0.623 (0.555–0.691) | 0.859 (0.825–0.890) |
| | Acoustic aggregates | LightGBM | 0.626 (0.556–0.699) | 0.864 (0.833–0.892) |
| | Acoustic aggregates, age and sex | Logistic regression | 0.624 (0.552–0.693) | 0.859 (0.825–0.890) |
| | Acoustic aggregates, age and sex | LightGBM | 0.621 (0.551–0.693) | 0.861 (0.829–0.890) |

Post hoc. Patient level: 736 children for the first two targets (majority-class rate 0.595) and 509 for pneumonia versus healthy controls (0.861). Out-of-fold probabilities from a patient-grouped five-fold cross-validation, averaged over five repetitions; intervals by bootstrap over children (1000 replicates). Logistic regression: standardized features, C = 0.1. LightGBM: 200 trees, learning rate 0.03, maximum depth 3. The unfitted mean predicted probabilities, which are counted among the 42 configurations, and the recording-level configurations are listed in the repository (results_clean/revision_requests/a2_posthoc_inventory.csv).

Median age was 5.0 years (interquartile range 3.4 to 7.4) in pneumonia, 4.2 (3.4 to 6.0) in bronchial disease, 4.9 (3.1 to 7.1) in healthy controls and 6.4 (4.0 to 9.8) in the other diagnoses of the normal/other group. Among children with pneumonia, the 40 with severe disease had a higher median fraction of events predicted adventitious than the 398 with non-severe disease (0.14 versus 0.07), but this fraction separated them only weakly (AUC 0.59, 95% CI 0.50 to 0.68); the mean predicted probability of pneumonia did no better (AUC 0.61, 95% CI 0.51 to 0.71).

## Operating point by event duration (post hoc)

Within tertiles of annotated event duration (short, up to 1272 ms; medium, up to 1936 ms; long, above 1936 ms) the AUC of both acoustic outcomes stayed at or above 0.902 (Section S5), but the decisions of the models shifted with duration (Table S6). Of the 3297 adventitious events, 2355 (71%) lay in the shortest tertile. At the model's own decision rule, sensitivity for an adventitious event fell from 0.885 in the shortest to 0.526 in the longest tertile while specificity rose from 0.866 to 0.977; for crackles, sensitivity fell to 0.267. With the threshold on the probability of an adventitious sound set within each tertile to reproduce the overall specificity (0.936), sensitivity in the longest tertile was 0.681, so part of the loss reflects the operating point and part a weaker ranking of long events. Discrimination is therefore not explained by duration, but the operating point depends on it, and a deployed model would need duration-aware thresholds or recalibration.

**Table S6 Sensitivity and specificity of the acoustic outcomes by tertile of annotated event duration (nested cross-validation)**

| Class | Tertile (events) | Sensitivity (95% CI) | Specificity (95% CI) | Sensitivity at overall specificity (95% CI) |
|---|---|---|---|---|
| Adventitious (screening) | Short (2355) | 0.885 (0.842–0.920) | 0.866 (0.842–0.889) | 0.797 (0.737–0.847) |
| | Medium (594) | 0.646 (0.573–0.715) | 0.944 (0.936–0.952) | 0.673 (0.607–0.744) |
| | Long (348) | 0.526 (0.414–0.627) | 0.977 (0.972–0.981) | 0.681 (0.593–0.759) |
| Crackles (sound pattern) | Short (1273) | 0.788 (0.735–0.839) | 0.866 (0.844–0.887) | 0.723 (0.633–0.802) |
| | Medium (412) | 0.592 (0.510–0.668) | 0.945 (0.936–0.952) | 0.675 (0.590–0.748) |
| | Long (161) | 0.267 (0.168–0.371) | 0.976 (0.971–0.981) | 0.559 (0.434–0.671) |
| Wheeze/rhonchi (sound pattern) | Short (1082) | 0.848 (0.770–0.904) | 0.866 (0.844–0.887) | 0.874 (0.820–0.913) |
| | Medium (182) | 0.637 (0.532–0.738) | 0.945 (0.936–0.952) | 0.758 (0.653–0.842) |
| | Long (187) | 0.684 (0.523–0.807) | 0.976 (0.971–0.981) | 0.802 (0.692–0.887) |

Not pre-specified. Full stacked model, pooled nested cross-validation (19693 events); the number of events of the class is given in parentheses. Sensitivity and specificity at the model's own decision (highest class probability). The last column uses, within each tertile, the threshold on the probability of any adventitious sound that reproduces the overall specificity of the model (0.936 for both outcomes); for the sound-pattern classes it counts detection as any adventitious class. Intervals by bootstrap over children (2000 replicates).

## Positive predictive value at other prevalences (post hoc)

With the observed sensitivity and specificity, the positive predictive value for an adventitious event would be 0.40 at a prevalence of 5% and 0.58 at 10%, against 0.72 at the observed 16.7%. For a child, under the post hoc rule of at least two events predicted adventitious (sensitivity 191

of 241, 0.79; specificity 388 of 495, 0.78), it would be 0.29 at 10% and 0.48 at 20%, against 0.64 at the observed 32.7%; negative predictive values were 0.96 or more for events and 0.89 or more for children.

### Second stage without class weighting (post hoc)

To test whether the class-balanced second stage caused the disease-group result, the pre-specified nested cross-validation was repeated with class weighting switched off; the folds, features and tuning protocol were unchanged, and the weighted full stack, re-run in the first fold, reproduced its original AUCs to four decimals. For screening and sound pattern, the event-level AUC rose slightly (0.960, 95% CI 0.951 to 0.968, and 0.959, 95% CI 0.951 to 0.966; 0.894 and 0.889 at patient level), accuracy exceeded the majority-class rate at event level (0.934 and 0.923 against 0.833) and at patient level (0.745 and 0.719 against 0.673), and the lower confidence bounds still exceeded the duration-only AUCs (0.777 and 0.761). For disease group, accuracy rose from 0.44 to 0.598 at event level and from 0.46 to 0.580 at patient level, still below the majority-class rates of 0.619 and 0.595, while balanced accuracy fell to 0.355 and 0.347 and the Matthews correlation coefficient was 0.079 and 0.062; the AUC was 0.599 (95% CI 0.560 to 0.640) and 0.603 (95% CI 0.570 to 0.637). The full stack still exceeded demographics (difference in AUC 0.084, 95% CI 0.043 to 0.125; unadjusted $p = 0.0005$, the smallest attainable value). No pre-specified wording rule changed.

### Child-level decision curve (post hoc)

For the derived child label (any annotated adventitious event; 241 of 736 children), the post hoc count rule (at least two events predicted adventitious) had a higher net benefit than treating all children from a threshold probability of about 0.11 upward (0.234 against 0.209 at 0.15; 0.197 against 0.039 at 0.30) and a higher net benefit than the pre-specified confidence-weighted rule at every threshold from 0.05 to 0.40. The pre-specified rule exceeded treating all children only above a threshold of about 0.26 (0.091 against 0.039 at 0.30). Both rules are binary, the prevalence is that of this cohort, the label is derived rather than a clinical diagnosis, and the count rule needs confirmation.

## S3 Hyper-parameters and software

### Table S7 Training configuration of the base models

| Setting | Value |
|---|---|
| Encoder | HeAR ViT-L (google/hear-pytorch, revision f791cd42), all weights frozen |
| Adapters | LoRA, rank 8, scaling 16, query and value projections of transformer blocks 19 to 24 |
| Pooling | 512-D pooled embedding + 4-head attention pooling over 96 patch tokens (1024-D) |
| Head | 1536 → 512 → 128 → classes, layer/batch normalization, GELU, dropout 0.3 |
| Trainable parameters | 4.2 million of 307 million (1.4%) |
| Loss | focal loss, gamma 2, inverse-frequency class weights |
| Optimizer | AdamW, learning rate 0.0002, weight decay 0.01, gradient clipping at 1.0 |
| Schedule | 2 warm-up epochs then cosine decay; at most 25 epochs; early stopping on validation macro-F1, patience 6 |
| Batch size / precision | 32 / bfloat16 autocast |
| Training augmentation | event position jitter up to 200 ms |
| Seed | 42 (Python, NumPy, PyTorch, data-loader workers) |

### Table S8 Search space of the second-stage LightGBM (Optuna, tree-structured Parzen estimator; 100 trials for the hold-out analysis, 50 per outer fold in the nested cross-validation; objective: validation macro-F1)

| Hyper-parameter | Range |
|---|---|
| Number of trees | 100 to 800 (early stopping after 20 rounds) |
| Learning rate | 0.01 to 0.3, log scale |
| Number of leaves | 15 to 127 |
| Maximum depth | 3 to 12 |
| Minimum samples per leaf | 5 to 100 |
| Row and column subsampling | 0.6 to 1.0 |
| L1 and L2 regularization | 1e-8 to 10, log scale |
| Class weights | balanced |

Software: Python 3.12, PyTorch 2.9 (CUDA 12.8), transformers 4.50, LightGBM, Optuna, SHAP, scikit-learn; one NVIDIA RTX 3090. A hold-out run takes about 6 h and the nested cross-validation about 18 h on a shared card.

## S4 TRIPOD+AI checklist

The checklist follows the layout of the TRIPOD+AI template, and item wording follows Collins et al., BMJ 2024;385:e078378. Locations are given as sections of the manuscript rather than page numbers, because page numbers change in production.

### Table S9 TRIPOD+AI checklist

| Section/Topic | Item | D; E[1] | Checklist item | Reported in |
|---|---|---|---|---|
| **TITLE** | | | | |
| Title | 1 | D; E | Identify the study as developing or evaluating the performance of a multivariable prediction model, the target population, and the outcome to be predicted | Title |
| **ABSTRACT** | | | | |
| Abstract | 2 | D; E | Provide a summary of objectives, study design, setting, participants, sample size, predictors, outcome, statistical analysis, results, and conclusions. | Abstract |
| **INTRODUCTION** | | | | |
| Background | 3a | D; E | Explain the healthcare context (including whether diagnostic or prognostic) and rationale for developing or evaluating the prediction model, including references to existing models | Background, paragraphs 1–2 |
| | 3b | D; E | Describe the target population and the intended purpose of the prediction model in the context of the care pathway, including its intended users (e.g., healthcare professionals, patients, public) | Background, paragraphs 2–4; Discussion (final paragraph) |
| | 3c | D; E | Describe any known health inequalities between sociodemographic groups | Not addressed in detail: the source database reports age and sex only; performance by age band and sex is given in Results, Acoustic event classification and Additional file 1, Section S5 |
| Objectives | 4 | D; E | Specify the study objectives, including whether the study describes the development or validation of a prediction model (or both) | Background, final paragraph (two objectives) |
| **METHODS** | | | | |
| Data | 5a | D; E | Describe the sources of data separately for the development and evaluation datasets (e.g., randomized trial, cohort, routine care or registry data), the rationale for using these data, and representativeness of the data | Methods, Design and data; Evaluation |
| | 5b | D; E | Specify the dates of the collected participant data, including start and end of participant accrual; and, if applicable, end of follow-up | Methods, Design and data (dates are those of the SPRSound releases; see the cited data descriptor) |
| Participants | 6a | D; E | Specify key elements of the study setting (e.g., primary care, secondary care, general population) including the number and location of centers | Methods, Design and data |
| | 6b | D; E | Describe the eligibility criteria for study participants | Methods, Design and data; Fig. 1 |
| | 6c | D; E | Give details of any treatments received, and how they were handled during model development or evaluation, if relevant | Not applicable: treatments are not recorded in the database |

| Section/Topic | Item | D; E[1] | Checklist item | Reported in |
|---|---|---|---|---|
| Data preparation | 7 | D; E | Describe any data pre-processing and quality checking, including whether this was similar across relevant sociodemographic groups | Methods, Design and data; Model |
| Outcome | 8a | D; E | Clearly define the outcome that is being predicted and the time horizon, including how and when assessed, the rationale for choosing this outcome, and whether the method of outcome assessment is consistent across sociodemographic groups | Methods, Outcomes |
| | 8b | D; E | If outcome assessment requires subjective interpretation, describe the qualifications and demographic characteristics of the outcome assessors | Methods, Design and data (annotation by pediatric physicians); Discussion (participants with conflicting diagnoses between source files) |
| | 8c | D; E | Report any actions to blind assessment of the outcome to be predicted | Not applicable: outcomes are pre-existing annotations of a public database |
| Predictors | 9a | D | Describe the choice of initial predictors (e.g., literature, previous models, all available predictors) and any pre-selection of predictors before model building | Methods, Model; Comparators and statistics |
| | 9b | D; E | Clearly define all predictors, including how and when they were measured (and any actions to blind assessment of predictors for the outcome and other predictors) | Methods, Model; Comparators and statistics (full detail in Additional file 1, Section S1) |
| | 9c | D; E | If predictor measurement requires subjective interpretation, describe the qualifications and demographic characteristics of the predictor assessors | Not applicable: predictors are audio, age, sex, auscultation site and annotated event duration |
| Sample size | 10 | D; E | Explain how the study size was arrived at (separately for development and evaluation) and justify that the study size was sufficient to answer the research question. Include details of any sample size calculation | No formal sample-size calculation: all eligible events of the public database were used. Precision is shown by patient-bootstrap intervals and between-fold ranges; the limited size of the hold-out is discussed (Results; Discussion (limitations)) |
| Missing data | 11 | D; E | Describe how missing data were handled. Provide reasons for omitting any data | Methods, Design and data (exclusions; no imputation) |
| Analytical methods | 12a | D | Describe how the data were used (e.g., for development and evaluation of model performance) in the analysis, including whether the data were partitioned, considering any sample size requirements | Methods, Evaluation |
| | 12b | D | Depending on the type of model, describe how predictors were handled in the analyses (functional form, rescaling, transformation, or any standardization). | Methods, Model |
| | 12c | D | Specify the type of model, rationale[2], all model-building steps, including any hyperparameter tuning, and method for internal validation | Methods, Model; Additional file 1, Sections S1 and S3 |
| | 12d | D; E | Describe if and how any heterogeneity in estimates of model parameter values and model performance was handled and quantified across clusters (e.g., hospitals, countries). See TRIPOD-Cluster for additional considerations[3] | Methods, Comparators and statistics (between-fold dispersion; exploratory subgroups). Single center and device, so no cluster-level heterogeneity could be assessed |

| Section/Topic | Item | D; E[1] | Checklist item | Reported in |
|---|---|---|---|---|
| | 12e | D; E | Specify all measures and plots used (and their rationale) to evaluate model performance (e.g., discrimination, calibration, clinical utility) and, if relevant, to compare multiple models | Methods, Comparators and statistics |
| | 12f | E | Describe any model updating (e.g., recalibration) arising from the model evaluation, either overall or for particular sociodemographic groups or settings | Not applicable: no model updating was performed; the need for recalibration is reported (Results, Acoustic event classification (Additional file 1, Section S2 for the full tables)) |
| | 12g | E | For model evaluation, describe how the model predictions were calculated (e.g., formula, code, object, application programming interface) | Methods, Model; Evaluation; Availability of data and materials (code) |
| Class imbalance | 13 | D; E | If class imbalance methods were used, state why and how this was done, and any subsequent methods to recalibrate the model or the model predictions | Methods, Model (focal loss with inverse-frequency class weights; class-balanced second stage); consequence for calibration in Results |
| Fairness | 14 | D; E | Describe any approaches that were used to address model fairness and their rationale | No fairness intervention was applied. Performance by age band, sex and auscultation site is reported (Additional file 1, Section S5); a demographics-only comparator quantifies reliance on age and sex (Results, Disease-group prediction) |
| Model output | 15 | D | Specify the output of the prediction model (e.g., probabilities, classification). Provide details and rationale for any classification and how the thresholds were identified | Methods, Model; Comparators and statistics (aggregation rules) |
| Training versus evaluation | 16 | D; E | Identify any differences between the development and evaluation data in healthcare setting, eligibility criteria, outcome, and predictors | Methods, Evaluation (development and evaluation data come from the same center, device and eligibility criteria) |
| Ethical approval | 17 | D; E | Name the institutional research board or ethics committee that approved the study and describe the participant-informed consent or the ethics committee waiver of informed consent | Declarations, Ethics approval and consent to participate |
| **OPEN SCIENCE** | | | | |
| Funding | 18a | D; E | Give the source of funding and the role of the funders for the present study | Declarations, Funding |
| Conflicts of interest | 18b | D; E | Declare any conflicts of interest and financial disclosures for all authors | Declarations, Competing interests |
| Protocol | 18c | D; E | Indicate where the study protocol can be accessed or state that a protocol was not prepared | Analysis plan written before the reported analyses, kept in the code repository; no separate protocol |
| Registration | 18d | D; E | Provide registration information for the study, including register name and registration number, or state that the study was not registered | Not registered |
| Data sharing | 18e | D; E | Provide details of the availability of the study data | Declarations, Availability of data and materials |
| Code sharing | 18f | D; E | Provide details of the availability of the analytical code[4] | Declarations, Availability of data and materials (public repository, MIT license) |

| Section/Topic | Item | D; E[1] | Checklist item | Reported in |
|---|---|---|---|---|
| **PATIENT & PUBLIC INVOLVEMENT** | | | | |
| Patient & Public Involvement | 19 | D; E | Provide details of any patient and public involvement during the design, conduct, reporting, interpretation, or dissemination of the study or state no involvement. | None; stated in Methods, Design and data |
| **RESULTS** | | | | |
| Participants | 20a | D; E | Describe the flow of participants through the study, including the number of participants with and without the outcome and, if applicable, a summary of the follow-up time. A diagram may be helpful. | Results, Cohort; Fig. 1 |
| | 20b | D; E | Report the characteristics overall and, where applicable, for each data source or setting, including the key dates, key predictors (including demographics), treatments received, sample size, number of outcome events, follow-up time, and amount of missing data. A table may be helpful. Report any differences across key demographic groups. | Results, Cohort; Table 1 |
| | 20c | E | For model evaluation, show a comparison with the development data of the distribution of important predictors (demographics, predictors, and outcome). | Table 1 (training versus test partition) |
| Model development | 21 | D; E | Specify the number of participants and outcome events in each analysis (e.g., for model development, hyperparameter tuning, model evaluation) | Table 1; Table 2 (units per analysis); Additional file 1, Table S1 (events per class) |
| Model specification | 22 | D | Provide details of the full prediction model (e.g., formula, code, object, application programming interface) to allow predictions in new individuals and to enable third-party evaluation and implementation, including any restrictions to access or re-use (e.g., freely available, proprietary)[5] | Methods, Model; code repository; trained adapter and head weights to be released (frozen encoder available from its publisher) |
| Model performance | 23a | D; E | Report model performance estimates with confidence intervals, including for any key subgroups (e.g., sociodemographic). Consider plots to aid presentation. | Results; Tables 2 and 3; Figs. 2–4; Additional file 1, Sections S5 and S6 |
| | 23b | D; E | If examined, report results of any heterogeneity in model performance across clusters. See TRIPOD Cluster for additional details[3]. | Results (between-fold ranges; subgroups); Additional file 1, Sections S5 and S6 |
| Model updating | 24 | E | Report the results from any model updating, including the updated model and subsequent performance | Not applicable |
| **DISCUSSION** | | | | |
| Interpretation | 25 | D; E | Give an overall interpretation of the main results, including issues of fairness in the context of the objectives and previous studies | Discussion (principal findings; why disease-group prediction failed) |
| Limitations | 26 | D; E | Discuss any limitations of the study (such as a non-representative sample, sample size, overfitting, missing data) and their effects on any biases, statistical uncertainty, and generalizability | Discussion (limitations) |
| Usability of the model in the context of current care | 27a | D | Describe how poor quality or unavailable input data (e.g., predictor values) should be assessed and handled when implementing the prediction model | Discussion (dependence on expert segmentation and on preprocessing) |

| Section/Topic | Item | D; E[1] | Checklist item | Reported in |
|---|---|---|---|---|
| | 27b | D | Specify whether users will be required to interact in the handling of the input data or use of the model, and what level of expertise is required of users | Discussion (events must be segmented by a clinician) |
| | 27c | D; E | Discuss any next steps for future research, with a specific view to applicability and generalizability of the model | Discussion (limitations; final paragraph: external, multi-device validation and event detection are required; no deployment claim) |

[1]D=items relevant only to the development of a prediction model; E=items relating solely to the evaluation of a prediction model; D; E=items applicable to both the development and evaluation of a prediction model [2]Separately for all model building approaches. [3]TRIPOD-Cluster is a checklist of reporting recommendations for studies developing or validating models that explicitly account for clustering or explore heterogeneity in model performance (e.g., at different hospitals or centers). Debray et al, BMJ 2023; 380: e071018 [DOI: 10.1136/bmj-2022-071018] [4]This relates to the analysis code, for example, any data cleaning, feature engineering, model building, evaluation. [5]This relates to the code to implement the model to get estimates of risk for a new individual.

## S5 Subgroup performance

Subgroup performance of the full stacked model in the nested cross-validation (primary analysis) by age band, sex, auscultation site and tertile of annotated event duration. Subgroup analyses are exploratory; event-duration tertiles were added to the planned subgroups (age band, sex and auscultation site) as a check for the duration shortcut. AUC is macro-averaged one-vs-rest for the three-class outcomes; sensitivity and specificity are given for the binary screening outcome only.

### Table S10 Subgroup performance of the full stacked model at event level (nested cross-validation)

| Outcome | Subgroup | Level | Patients | Events | AUC (95% CI) | Sensitivity | Specificity |
|---|---|---|---|---|---|---|---|
| Screening | Age band | <3 y | 141 | 5068 | 0.950 (0.929–0.967) | 0.891 | 0.882 |
| | | 3 to <6 y | 322 | 8284 | 0.938 (0.919–0.954) | 0.737 | 0.944 |
| | | 6 to <12 y | 244 | 5936 | 0.928 (0.896–0.949) | 0.653 | 0.955 |
| | | ≥12 y | 31 | 405 | 0.959 (0.916–0.998) | 0.800 | 0.989 |
| | Sex | Female | 362 | 9180 | 0.943 (0.926–0.957) | 0.769 | 0.939 |
| | | Male | 374 | 10513 | 0.951 (0.934–0.966) | 0.828 | 0.934 |
| | Auscultation site | Left posterior (p1) | 590 | 4981 | 0.957 (0.942–0.968) | 0.829 | 0.940 |
| | | Left lateral (p2) | 612 | 5329 | 0.934 (0.911–0.953) | 0.790 | 0.926 |
| | | Right posterior (p3) | 568 | 4562 | 0.951 (0.933–0.965) | 0.798 | 0.938 |
| | | Right lateral (p4) | 573 | 4821 | 0.953 (0.938–0.965) | 0.806 | 0.942 |
| | Event-duration tertile | Short | 541 | 6570 | 0.945 (0.928–0.958) | 0.885 | 0.866 |
| | | Medium | 647 | 6567 | 0.926 (0.908–0.942) | 0.646 | 0.944 |
| | | Long | 614 | 6556 | 0.902 (0.864–0.930) | 0.526 | 0.977 |
| Sound pattern | Age band | <3 y | 141 | 5068 | 0.946 (0.926–0.961) | | |
| | | 3 to <6 y | 322 | 8284 | 0.944 (0.927–0.957) | | |
| | | 6 to <12 y | 244 | 5936 | 0.937 (0.906–0.956) | | |
| | | ≥12 y | 31 | 405 | 0.965 (0.922–0.997) | | |
| | Sex | Female | 362 | 9180 | 0.947 (0.933–0.959) | | |
| | | Male | 374 | 10513 | 0.951 (0.936–0.963) | | |
| | Auscultation site | Left posterior (p1) | 590 | 4981 | 0.956 (0.942–0.966) | | |
| | | Left lateral (p2) | 612 | 5329 | 0.936 (0.913–0.954) | | |
| | | Right posterior (p3) | 568 | 4562 | 0.953 (0.939–0.964) | | |
| | | Right lateral (p4) | 573 | 4821 | 0.956 (0.943–0.966) | | |
| | Event-duration tertile | Short | 541 | 6570 | 0.947 (0.934–0.958) | | |
| | | Medium | 647 | 6567 | 0.932 (0.914–0.947) | | |
| | | Long | 614 | 6556 | 0.908 (0.875–0.933) | | |
| Disease group | Age band | <3 y | 141 | 5068 | 0.540 (0.450–0.635) | | |
| | | 3 to <6 y | 322 | 8284 | 0.645 (0.593–0.695) | | |
| | | 6 to <12 y | 244 | 5936 | 0.504 (0.431–0.579) | | |
| | | ≥12 y | 31 | 405 | 0.390 (0.304–0.506) | | |
| | Sex | Female | 362 | 9180 | 0.538 (0.480–0.599) | | |
| | | Male | 374 | 10513 | 0.601 (0.541–0.655) | | |
| | Auscultation site | Left posterior (p1) | 590 | 4981 | 0.574 (0.531–0.617) | | |
| | | Left lateral (p2) | 612 | 5329 | 0.596 (0.551–0.642) | | |
| | | Right posterior (p3) | 568 | 4562 | 0.579 (0.532–0.628) | | |

| Outcome | Subgroup | Level | Patients | Events | AUC (95% CI) | Sensitivity | Specificity |
|---|---|---|---|---|---|---|---|
| | | Right lateral (p4) | 573 | 4821 | 0.583 (0.539–0.628) | | |
| | Event-duration tertile | Short | 541 | 6570 | 0.577 (0.507–0.656) | | |
| | | Medium | 647 | 6567 | 0.584 (0.541–0.626) | | |
| | | Long | 614 | 6556 | 0.536 (0.491–0.583) | | |

Intervals are percentile bootstrap intervals obtained by resampling patients.

## S6 Further metrics of the primary analysis

Event-level metrics of every comparator model, metrics after aggregation of event probabilities to recordings and patients, and metrics of the full stacked model in each outer fold, all from the nested cross-validation (pooled out-of-fold predictions of the five outer folds). Point estimates are shown above their 95% confidence intervals, which are percentile bootstrap intervals obtained by resampling patients. AUC and the area under the precision–recall curve (AUPRC) are macro-averaged one-vs-rest for three-class outcomes; ECE, expected calibration error (ten bins).

### Table S11 Event-level performance of every model in the nested cross-validation

| Outcome | Model | Majority-class accuracy | Accuracy | Balanced accuracy | Macro-F1 | MCC | AUC | AUPRC | Brier score | ECE |
|---|---|---|---|---|---|---|---|---|---|---|
| Screening | Event duration only | 0.833 | 0.777 | 0.730 | 0.677 | 0.384 | 0.779 | 0.448 | 0.357 | 0.084 |
| | | | (0.756–0.797) | (0.695–0.761) | (0.643–0.708) | (0.319–0.444) | (0.740–0.813) | (0.364–0.528) | (0.338–0.378) | (0.069–0.099) |
| | Demographics only | 0.833 | 0.730 | 0.644 | 0.608 | 0.241 | 0.653 | 0.271 | 0.405 | 0.089 |
| | | | (0.693–0.765) | (0.594–0.692) | (0.565–0.650) | (0.154–0.323) | (0.578–0.715) | (0.201–0.370) | (0.373–0.442) | (0.053–0.124) |
| | Duration and demographics | 0.833 | 0.776 | 0.725 | 0.674 | 0.376 | 0.771 | 0.397 | 0.339 | 0.052 |
| | | | (0.749–0.802) | (0.683–0.763) | (0.634–0.710) | (0.301–0.445) | (0.720–0.814) | (0.307–0.521) | (0.308–0.376) | (0.034–0.076) |
| | Own-task base model | 0.833 | 0.897 | 0.894 | 0.839 | 0.695 | 0.960 | 0.875 | 0.159 | 0.018 |
| | | | (0.888–0.905) | (0.880–0.907) | (0.817–0.859) | (0.656–0.731) | (0.951–0.968) | (0.835–0.906) | (0.147–0.170) | (0.013–0.025) |
| | All acoustic probabilities | 0.833 | 0.914 | 0.882 | 0.856 | 0.715 | 0.952 | 0.851 | 0.131 | 0.023 |
| | | | (0.905–0.922) | (0.864–0.898) | (0.834–0.874) | (0.674–0.751) | (0.941–0.961) | (0.805–0.888) | (0.120–0.144) | (0.017–0.029) |
| | Full stack | 0.833 | 0.914 | 0.870 | 0.853 | 0.709 | 0.948 | 0.846 | 0.130 | 0.020 |
| | | | (0.906–0.923) | (0.848–0.890) | (0.831–0.873) | (0.665–0.747) | (0.936–0.958) | (0.798–0.885) | (0.117–0.142) | (0.014–0.027) |
| Sound pattern | Event duration only | 0.833 | 0.722 | 0.579 | 0.525 | 0.324 | 0.748 | 0.483 | 0.484 | 0.170 |
| | | | (0.696–0.746) | (0.540–0.615) | (0.490–0.554) | (0.270–0.372) | (0.715–0.778) | (0.450–0.518) | (0.465–0.504) | (0.146–0.193) |
| | Demographics only | 0.833 | 0.664 | 0.463 | 0.406 | 0.176 | 0.619 | 0.385 | 0.530 | 0.127 |
| | | | (0.621–0.704) | (0.412–0.509) | (0.369–0.441) | (0.110–0.238) | (0.560–0.669) | (0.362–0.414) | (0.501–0.561) | (0.095–0.168) |
| | Duration and demographics | 0.833 | 0.698 | 0.532 | 0.474 | 0.272 | 0.734 | 0.454 | 0.428 | 0.040 |
| | | | (0.664–0.729) | (0.490–0.573) | (0.441–0.506) | (0.216–0.324) | (0.693–0.769) | (0.425–0.486) | (0.394–0.466) | (0.025–0.063) |
| | Own-task base model | 0.833 | 0.849 | 0.815 | 0.730 | 0.605 | 0.936 | 0.810 | 0.221 | 0.051 |
| | | | (0.839–0.859) | (0.789–0.836) | (0.698–0.758) | (0.562–0.645) | (0.925–0.947) | (0.769–0.845) | (0.206–0.236) | (0.044–0.059) |
| | All acoustic probabilities | 0.833 | 0.902 | 0.814 | 0.784 | 0.691 | 0.949 | 0.833 | 0.153 | 0.033 |
| | | | (0.891–0.912) | (0.789–0.838) | (0.756–0.807) | (0.656–0.724) | (0.940–0.957) | (0.800–0.862) | (0.138–0.171) | (0.026–0.041) |
| | Full stack | 0.833 | 0.904 | 0.812 | 0.788 | 0.691 | 0.950 | 0.839 | 0.148 | 0.027 |
| | | | (0.893–0.914) | (0.783–0.838) | (0.759–0.812) | (0.651–0.727) | (0.940–0.958) | (0.805–0.868) | (0.133–0.163) | (0.021–0.035) |
| Disease group | Event duration only | 0.619 | 0.427 | 0.375 | 0.363 | 0.075 | 0.568 | 0.372 | 0.654 | 0.030 |
| | | | (0.399–0.456) | (0.352–0.398) | (0.341–0.383) | (0.040–0.110) | (0.543–0.593) | (0.356–0.392) | (0.645–0.662) | (0.009–0.059) |

| Outcome | Model | Majority-class accuracy | Accuracy | Balanced accuracy | Macro-F1 | MCC | AUC | AUPRC | Brier score | ECE |
|---|---|---|---|---|---|---|---|---|---|---|
| | Demographics only | 0.619 | 0.382 (0.331–0.431) | 0.351 (0.301–0.403) | 0.337 (0.290–0.381) | 0.036 (−0.033 to 0.110) | 0.517 (0.474–0.563) | 0.346 (0.324–0.378) | 0.669 (0.653–0.685) | 0.045 (0.026–0.101) |
| | Duration and demographics | 0.619 | 0.423 (0.381–0.468) | 0.372 (0.332–0.417) | 0.362 (0.325–0.400) | 0.082 (0.025–0.141) | 0.574 (0.535–0.613) | 0.379 (0.355–0.411) | 0.644 (0.625–0.663) | 0.032 (0.014–0.075) |
| | Own-task base model | 0.619 | 0.407 (0.377–0.438) | 0.419 (0.394–0.444) | 0.377 (0.351–0.403) | 0.127 (0.089–0.166) | 0.593 (0.565–0.621) | 0.386 (0.369–0.407) | 0.651 (0.640–0.662) | 0.011 (0.007–0.040) |
| | All acoustic probabilities | 0.619 | 0.427 (0.397–0.458) | 0.415 (0.391–0.440) | 0.384 (0.360–0.409) | 0.122 (0.085–0.160) | 0.594 (0.566–0.623) | 0.389 (0.368–0.412) | 0.640 (0.619–0.660) | 0.031 (0.012–0.059) |
| | Full stack | 0.619 | 0.438 (0.395–0.479) | 0.391 (0.349–0.434) | 0.377 (0.338–0.414) | 0.104 (0.047–0.161) | 0.583 (0.543–0.624) | 0.382 (0.356–0.416) | 0.642 (0.616–0.669) | 0.055 (0.024–0.097) |

Model rows correspond to the comparator ladder (Table 3); all models use the same second-stage learner and tuning protocol. MCC, Matthews correlation coefficient.

## Table S12 Performance after aggregation to recordings and patients (nested cross-validation)

| Outcome | Level | Rule | Units | Majority-class accuracy | Accuracy | Balanced accuracy | Macro-F1 | MCC | AUC | AUPRC | Brier score | ECE |
|---|---|---|---|---|---|---|---|---|---|---|---|---|
| Screening | Recording | Confidence-weighted | 4985 | 0.779 | 0.907 (0.894–0.920) | 0.820 (0.794–0.844) | 0.851 (0.829–0.871) | 0.716 (0.675–0.753) | 0.938 (0.924–0.950) | 0.862 (0.822–0.894) | 0.141 (0.124–0.158) | 0.011 (0.007–0.020) |
| | Recording | Mean | 4985 | 0.779 | 0.906 (0.894–0.919) | 0.817 (0.791–0.842) | 0.849 (0.828–0.869) | 0.712 (0.673–0.749) | 0.937 (0.923–0.949) | 0.860 (0.819–0.892) | 0.143 (0.127–0.159) | 0.012 (0.009–0.023) |
| | Patient | Confidence-weighted | 736 | 0.673 | 0.761 (0.730–0.792) | 0.638 (0.609–0.668) | 0.642 (0.600–0.683) | 0.439 (0.379–0.496) | 0.872 (0.842–0.901) | 0.816 (0.772–0.858) | 0.342 (0.298–0.384) | 0.123 (0.096–0.152) |
| | Patient | Mean | 736 | 0.673 | 0.762 (0.732–0.793) | 0.640 (0.612–0.669) | 0.645 (0.603–0.685) | 0.443 (0.383–0.500) | 0.869 (0.839–0.897) | 0.813 (0.769–0.854) | 0.335 (0.293–0.375) | 0.111 (0.086–0.141) |
| Sound pattern | Recording | Confidence-weighted | 4985 | 0.779 | 0.884 (0.869–0.899) | 0.698 (0.666–0.727) | 0.746 (0.719–0.770) | 0.657 (0.621–0.690) | 0.942 (0.931–0.952) | 0.832 (0.803–0.856) | 0.166 (0.147–0.186) | 0.009 (0.006–0.020) |
| | Recording | Mean | 4985 | 0.779 | 0.886 (0.871–0.901) | 0.702 (0.671–0.730) | 0.751 (0.724–0.775) | 0.663 (0.627–0.695) | 0.942 (0.930–0.951) | 0.830 (0.801–0.855) | 0.168 (0.148–0.187) | 0.015 (0.010–0.025) |
| | Patient | Confidence-weighted | 736 | 0.673 | 0.742 (0.709–0.774) | 0.482 (0.448–0.520) | 0.517 (0.466–0.567) | 0.400 (0.341–0.460) | 0.880 (0.857–0.903) | 0.747 (0.709–0.787) | 0.378 (0.333–0.422) | 0.134 (0.108–0.166) |
| | Patient | Mean | 736 | 0.673 | 0.742 (0.711–0.774) | 0.482 (0.447–0.519) | 0.517 (0.466–0.565) | 0.400 (0.341–0.457) | 0.879 (0.855–0.901) | 0.746 (0.709–0.786) | 0.371 (0.327–0.414) | 0.123 (0.096–0.156) |

| Outcome | Level | Rule | Units | Majority-class accuracy | Accuracy | Balanced accuracy | Macro-F1 | MCC | AUC | AUPRC | Brier score | ECE |
|---|---|---|---|---|---|---|---|---|---|---|---|---|
| Disease group | Recording | Confidence-weighted | 4985 | 0.625 | 0.455 (0.414–0.497) | 0.399 (0.359–0.441) | 0.386 (0.350–0.423) | 0.120 (0.067–0.173) | 0.600 (0.562–0.639) | 0.397 (0.371–0.428) | 0.629 (0.604–0.654) | 0.032 (0.017–0.075) |
| | Recording | Mean | 4985 | 0.625 | 0.455 (0.413–0.497) | 0.400 (0.360–0.442) | 0.387 (0.351–0.425) | 0.122 (0.069–0.176) | 0.600 (0.562–0.639) | 0.396 (0.371–0.428) | 0.630 (0.605–0.655) | 0.029 (0.017–0.072) |
| | Patient | Confidence-weighted | 736 | 0.595 | 0.457 (0.421–0.493) | 0.413 (0.375–0.453) | 0.401 (0.367–0.436) | 0.138 (0.087–0.192) | 0.602 (0.569–0.636) | 0.409 (0.386–0.443) | 0.633 (0.614–0.652) | 0.048 (0.025–0.085) |
| | Patient | Mean | 736 | 0.595 | 0.450 (0.414–0.488) | 0.409 (0.371–0.449) | 0.396 (0.362–0.432) | 0.131 (0.080–0.185) | 0.600 (0.567–0.635) | 0.408 (0.385–0.442) | 0.635 (0.616–0.653) | 0.050 (0.027–0.087) |

Confidence-weighted averaging was pre-specified; simple averaging (mean) is a sensitivity analysis. Recording- and patient-level labels are derived for screening and sound pattern (“any adventitious event”; most frequent adventitious pattern) and are the recording diagnosis for disease group.

## Table S13 Event-level performance of the full stacked model in each outer fold

| Outcome | Outer fold | Accuracy | Balanced accuracy | Macro-F1 | MCC | AUC | AUPRC | Brier score | ECE |
|---|---|---|---|---|---|---|---|---|---|
| Screening | Fold 1 | 0.912 | 0.874 | 0.847 | 0.698 | 0.953 | 0.870 | 0.130 | 0.007 |
| | Fold 2 | 0.907 | 0.899 | 0.838 | 0.693 | 0.950 | 0.844 | 0.147 | 0.037 |
| | Fold 3 | 0.916 | 0.751 | 0.770 | 0.544 | 0.908 | 0.657 | 0.127 | 0.025 |
| | Fold 4 | 0.925 | 0.878 | 0.877 | 0.754 | 0.958 | 0.880 | 0.115 | 0.017 |
| | Fold 5 | 0.910 | 0.891 | 0.876 | 0.754 | 0.961 | 0.903 | 0.131 | 0.017 |
| Sound pattern | Fold 1 | 0.896 | 0.814 | 0.756 | 0.663 | 0.950 | 0.842 | 0.154 | 0.013 |
| | Fold 2 | 0.904 | 0.838 | 0.773 | 0.682 | 0.947 | 0.832 | 0.157 | 0.048 |
| | Fold 3 | 0.912 | 0.666 | 0.691 | 0.546 | 0.917 | 0.727 | 0.136 | 0.031 |
| | Fold 4 | 0.916 | 0.853 | 0.840 | 0.744 | 0.967 | 0.895 | 0.128 | 0.025 |
| | Fold 5 | 0.891 | 0.815 | 0.789 | 0.723 | 0.955 | 0.841 | 0.164 | 0.023 |
| Disease group | Fold 1 | 0.531 | 0.477 | 0.447 | 0.220 | 0.677 | 0.459 | 0.582 | 0.080 |
| | Fold 2 | 0.510 | 0.470 | 0.460 | 0.200 | 0.636 | 0.426 | 0.622 | 0.052 |
| | Fold 3 | 0.449 | 0.395 | 0.378 | 0.106 | 0.557 | 0.370 | 0.685 | 0.133 |
| | Fold 4 | 0.417 | 0.383 | 0.361 | 0.088 | 0.609 | 0.393 | 0.618 | 0.080 |
| | Fold 5 | 0.292 | 0.242 | 0.245 | −0.072 | 0.477 | 0.335 | 0.712 | 0.192 |

# S7 Benchmark of second-stage learners

Benchmark of nine second-stage learners fitted to the same features, on the locked hold-out. An interval is available for the AUC only.

### Table S14 Benchmark of nine second-stage learners on the locked hold-out

| Outcome | Learner | Accuracy | Macro-F1 | MCC | AUC (95% CI) | Brier score |
|---|---|---|---|---|---|---|
| Screening | LightGBM | 0.939 | 0.869 | 0.738 | 0.963 (0.932–0.978) | 0.089 |
| | Random Forest | 0.944 | 0.879 | 0.759 | 0.962 (0.932–0.979) | 0.084 |
| | Extra Trees | 0.945 | 0.882 | 0.765 | 0.965 (0.936–0.979) | 0.084 |
| | Hist-GBDT | 0.940 | 0.873 | 0.746 | 0.964 (0.933–0.979) | 0.086 |
| | MLP (Neural Net) | 0.943 | 0.879 | 0.758 | 0.965 (0.935–0.978) | 0.086 |
| | Logistic Regression | 0.943 | 0.882 | 0.764 | 0.968 (0.941–0.981) | 0.083 |
| | XGBoost | 0.939 | 0.868 | 0.737 | 0.963 (0.930–0.979) | 0.086 |
| | CatBoost | 0.943 | 0.878 | 0.757 | 0.965 (0.933–0.980) | 0.085 |
| | Voting Ensemble | 0.941 | 0.873 | 0.746 | 0.964 (0.933–0.980) | 0.084 |
| Sound pattern | LightGBM | 0.933 | 0.802 | 0.721 | 0.962 (0.939–0.975) | 0.101 |
| | Random Forest | 0.938 | 0.811 | 0.740 | 0.962 (0.936–0.977) | 0.095 |
| | Extra Trees | 0.938 | 0.815 | 0.741 | 0.967 (0.947–0.978) | 0.094 |
| | Hist-GBDT | 0.934 | 0.808 | 0.732 | 0.966 (0.945–0.977) | 0.096 |
| | MLP (Neural Net) | 0.935 | 0.820 | 0.741 | 0.963 (0.938–0.976) | 0.098 |
| | Logistic Regression | 0.936 | 0.813 | 0.737 | 0.967 (0.947–0.978) | 0.094 |
| | XGBoost | 0.934 | 0.803 | 0.727 | 0.965 (0.944–0.978) | 0.097 |
| | CatBoost | 0.935 | 0.805 | 0.730 | 0.964 (0.940–0.977) | 0.096 |
| | Voting Ensemble | 0.935 | 0.805 | 0.728 | 0.964 (0.941–0.978) | 0.095 |
| Disease group | LightGBM | 0.655 | 0.384 | 0.211 | 0.678 (0.561–0.766) | 0.497 |
| | Random Forest | 0.691 | 0.355 | 0.124 | 0.737 (0.602–0.818) | 0.414 |
| | Extra Trees | 0.701 | 0.294 | 0.013 | 0.716 (0.601–0.793) | 0.426 |
| | Hist-GBDT | 0.673 | 0.398 | 0.231 | 0.677 (0.556–0.768) | 0.472 |
| | MLP (Neural Net) | 0.682 | 0.462 | 0.223 | 0.686 (0.538–0.776) | 0.445 |
| | Logistic Regression | 0.719 | 0.286 | 0.042 | 0.692 (0.584–0.776) | 0.422 |
| | XGBoost | 0.682 | 0.390 | 0.229 | 0.697 (0.565–0.785) | 0.429 |
| | CatBoost | 0.695 | 0.361 | 0.140 | 0.722 (0.571–0.812) | 0.406 |
| | Voting Ensemble | 0.685 | 0.379 | 0.227 | 0.706 (0.578–0.790) | 0.425 |

MCC, Matthews correlation coefficient.

## S8 SHAP attribution of the second-stage learners

SHAP values were computed with TreeExplainer for the full-stack LightGBM models of the locked hold-out (released encoder) on its test events. Demographic variables (age, sex, auscultation site) carried 2.8% of the mean absolute attribution for screening, 9.5% for sound pattern and 55.5% for disease group.

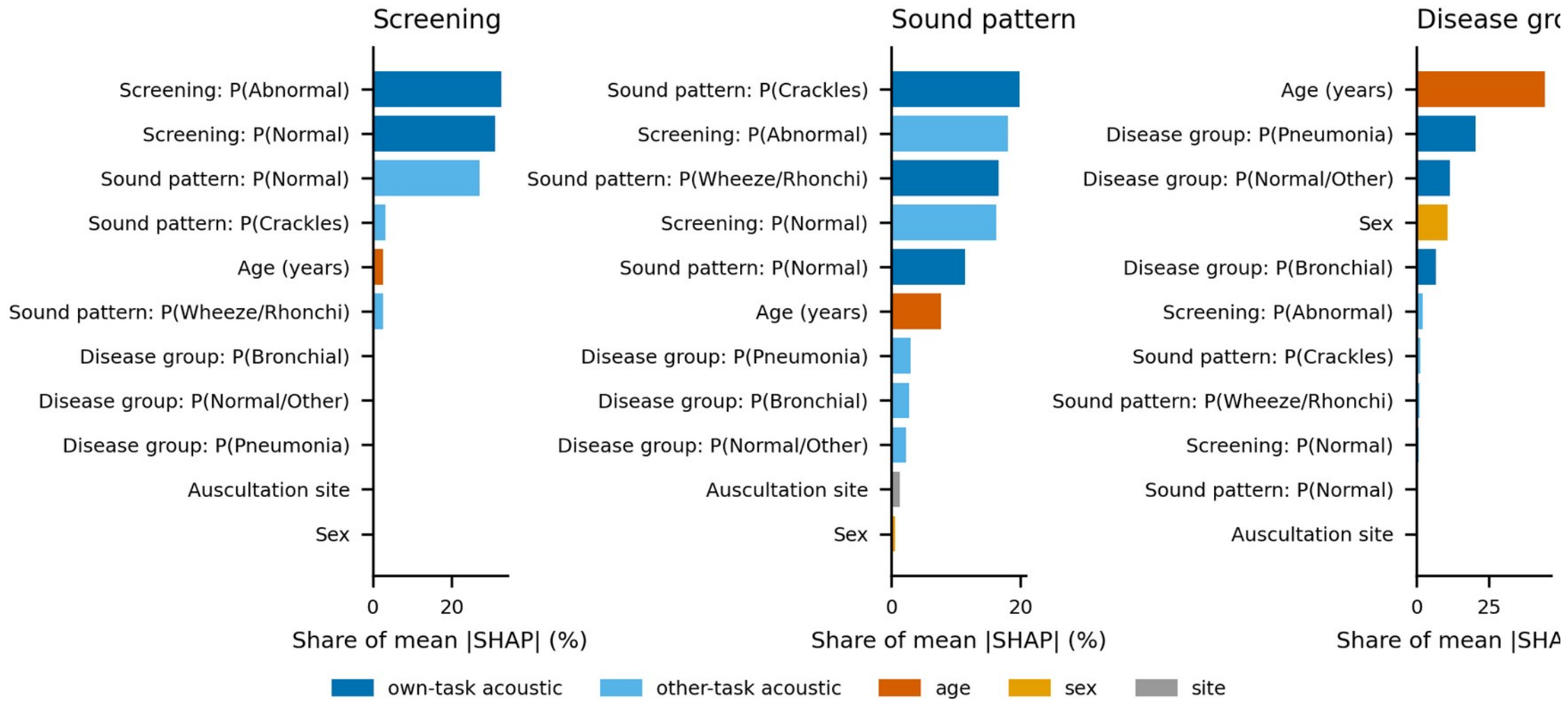

**Figure S1.** Share of mean absolute SHAP attribution by feature, colored by feature group, for the three outcomes.

### Table S15 Mean absolute SHAP value and share of attribution per feature

| Outcome | Feature | Group | Mean \|SHAP\| | Share (%) |
|---|---|---|---:|---:|
| Screening | Sound pattern: P(Normal) | other-task acoustic | 0.280 | 27.0 |
| Screening | Sound pattern: P(Crackles) | other-task acoustic | 0.033 | 3.2 |
| Screening | Sound pattern: P(Wheeze/Rhonchi) | other-task acoustic | 0.027 | 2.6 |
| Screening | Screening: P(Normal) | own-task acoustic | 0.320 | 30.9 |
| Screening | Screening: P(Abnormal) | own-task acoustic | 0.336 | 32.5 |
| Screening | Disease group: P(Pneumonia) | other-task acoustic | 0.003 | 0.2 |
| Screening | Disease group: P(Bronchial) | other-task acoustic | 0.005 | 0.5 |
| Screening | Disease group: P(Normal/Other) | other-task acoustic | 0.003 | 0.3 |
| Screening | Age (years) | age | 0.028 | 2.7 |
| Screening | Sex | sex | 0.000 | 0.0 |
| Screening | Auscultation site | site | 0.002 | 0.2 |
| Sound pattern | Sound pattern: P(Normal) | own-task acoustic | 0.180 | 11.5 |
| Sound pattern | Sound pattern: P(Crackles) | own-task acoustic | 0.312 | 19.9 |
| Sound pattern | Sound pattern: P(Wheeze/Rhonchi) | own-task acoustic | 0.261 | 16.7 |
| Sound pattern | Screening: P(Normal) | other-task acoustic | 0.254 | 16.2 |

| Outcome | Feature | Group | Mean \|SHAP\| | Share (%) |
|---|---|---|---|---|
| Sound pattern | Screening: P(Abnormal) | other-task acoustic | 0.283 | 18.1 |
| Sound pattern | Disease group: P(Pneumonia) | other-task acoustic | 0.047 | 3.0 |
| Sound pattern | Disease group: P(Bronchial) | other-task acoustic | 0.043 | 2.8 |
| Sound pattern | Disease group: P(Normal/Other) | other-task acoustic | 0.036 | 2.3 |
| Sound pattern | Age (years) | age | 0.120 | 7.7 |
| Sound pattern | Sex | sex | 0.009 | 0.5 |
| Sound pattern | Auscultation site | site | 0.020 | 1.3 |
| Disease group | Sound pattern: P(Normal) | other-task acoustic | 0.001 | 0.1 |
| Disease group | Sound pattern: P(Crackles) | other-task acoustic | 0.008 | 1.4 |
| Disease group | Sound pattern: P(Wheeze/Rhonchi) | other-task acoustic | 0.006 | 1.1 |
| Disease group | Screening: P(Normal) | other-task acoustic | 0.005 | 0.9 |
| Disease group | Screening: P(Abnormal) | other-task acoustic | 0.012 | 2.1 |
| Disease group | Disease group: P(Pneumonia) | own-task acoustic | 0.112 | 20.4 |
| Disease group | Disease group: P(Bronchial) | own-task acoustic | 0.037 | 6.7 |
| Disease group | Disease group: P(Normal/Other) | own-task acoustic | 0.064 | 11.7 |
| Disease group | Age (years) | age | 0.245 | 44.5 |
| Disease group | Sex | sex | 0.060 | 10.9 |
| Disease group | Auscultation site | site | 0.000 | 0.0 |

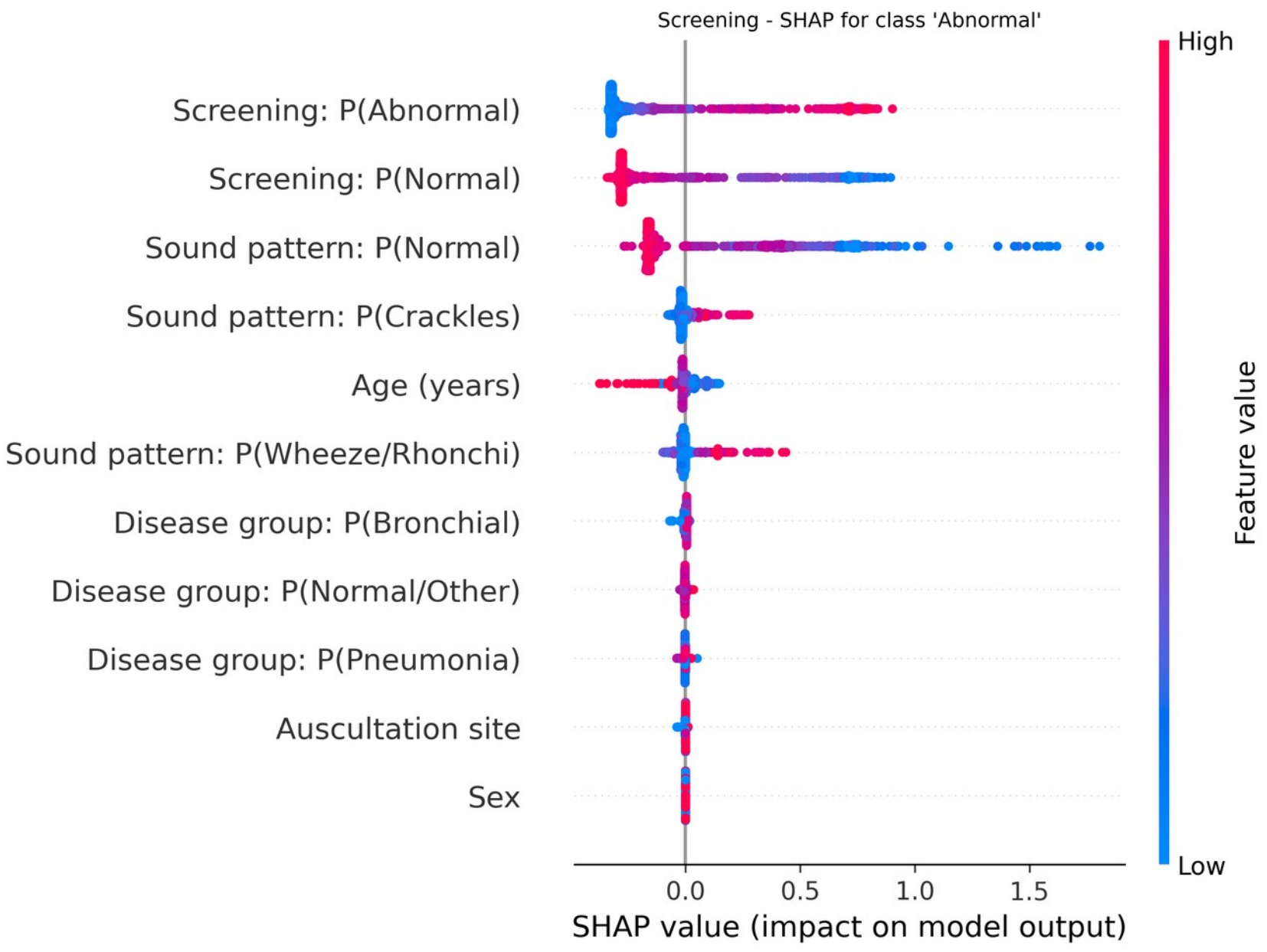

**Figure S2.** SHAP summary plot, screening (class: adventitious).

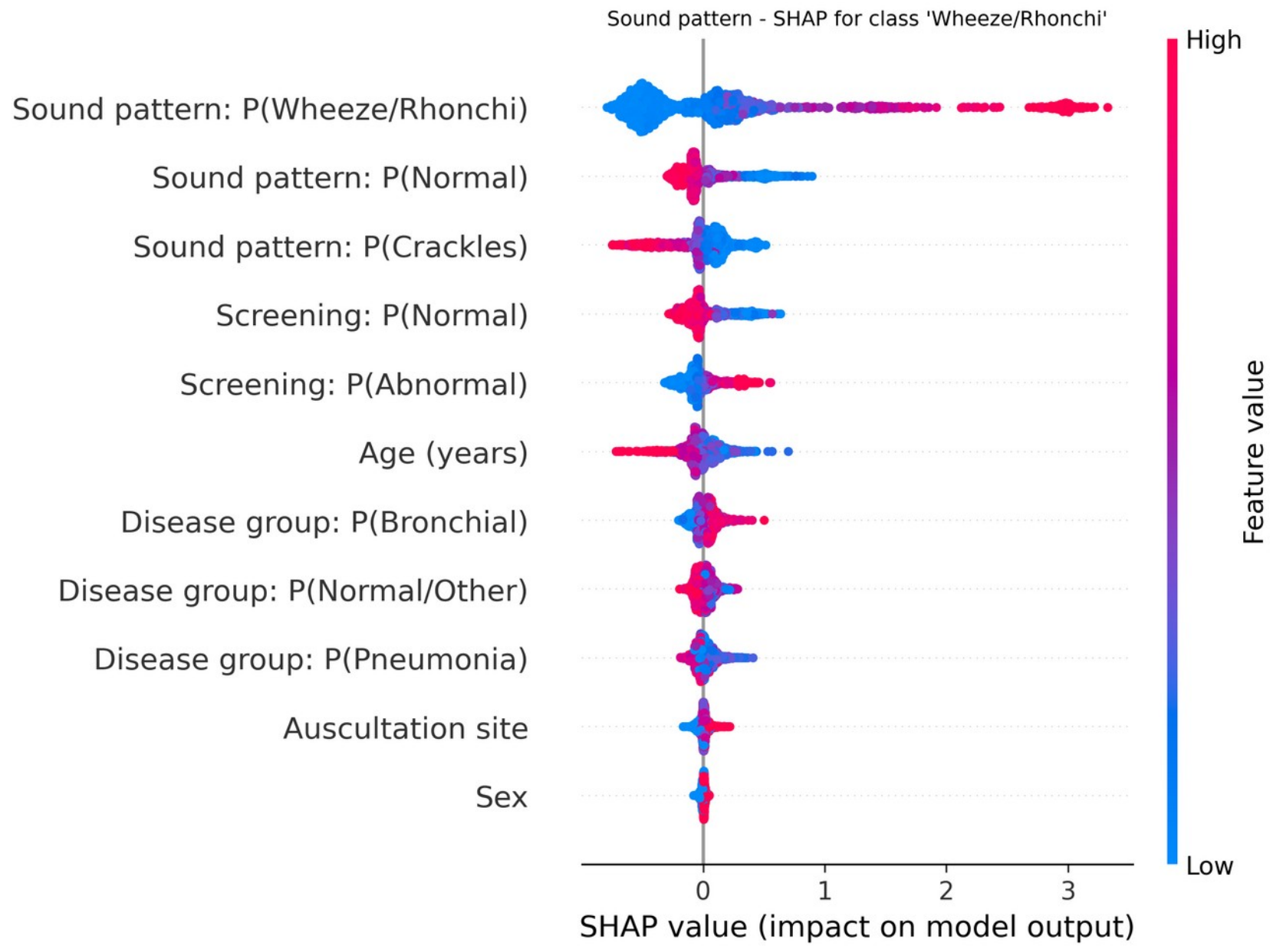

**Figure S3.** SHAP summary plot, sound pattern (class: wheeze/rhonchi).

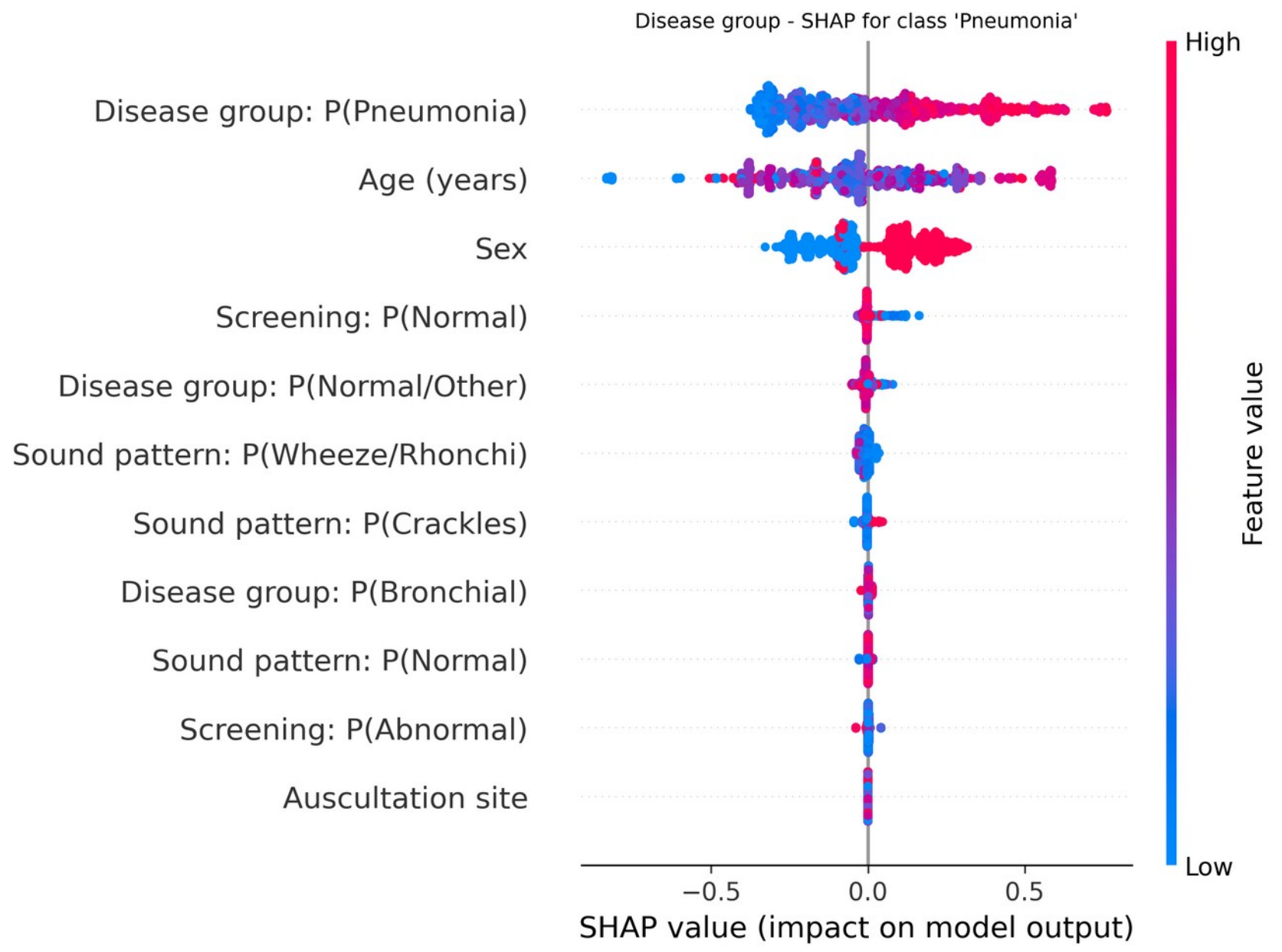

**Figure S4.** SHAP summary plot, disease group (class: pneumonia).

## S9 Model-derived attribution: further maps and faithfulness checks

Events were sampled from the locked test partition by a fixed seeded rule (up to 150 per true class), not selected by hand: 300 events for screening (266 correctly classified) and 414 for sound pattern (337 correctly classified).

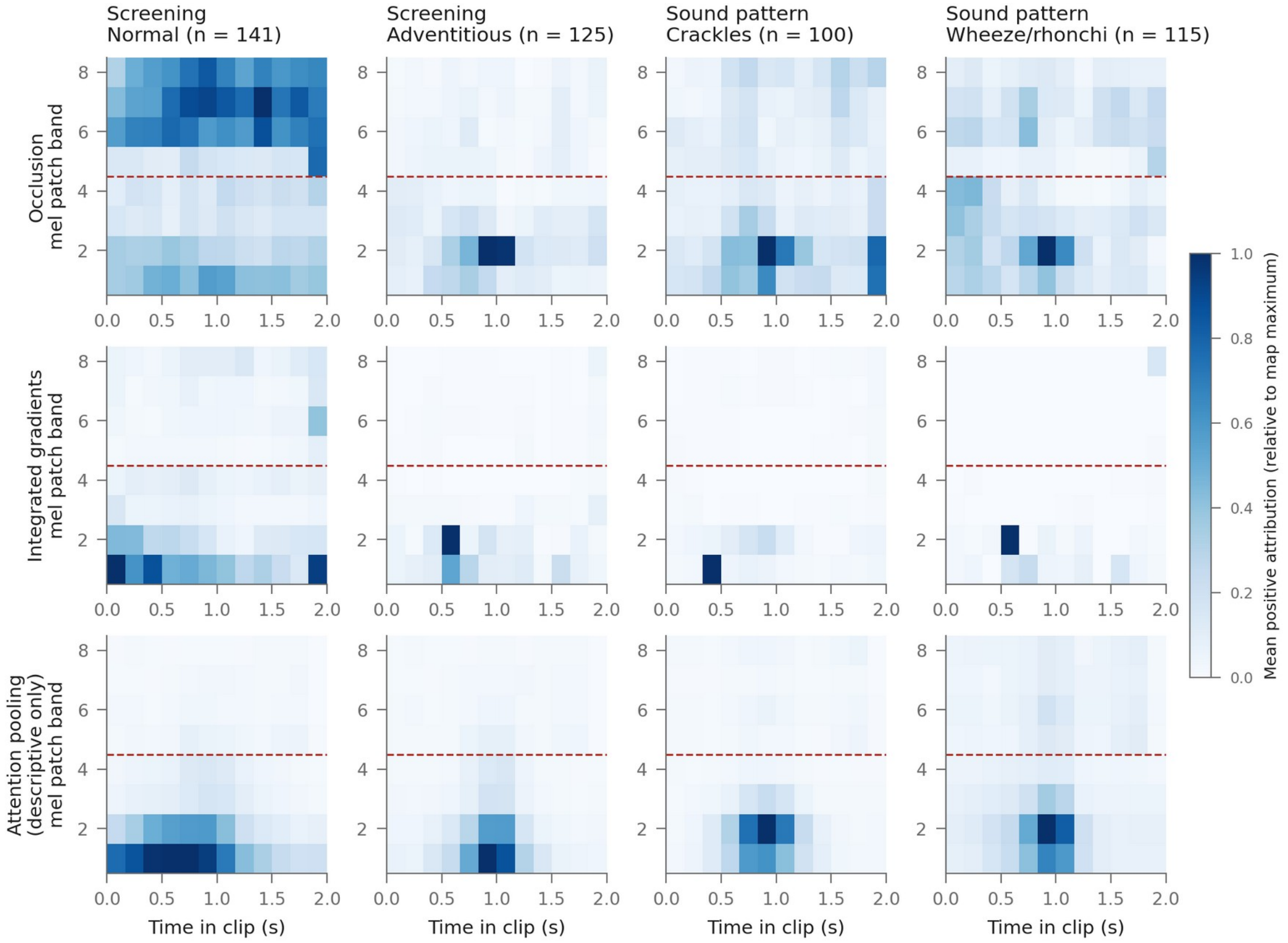

**Figure S5.** Class-mean attribution on the 12 x 8 (time x mel band) patch grid of the encoder input for correctly classified events, by three methods. The dashed line marks the cut-off of the band-pass filter applied during preprocessing; bands above it contain only attenuated residual energy. Occlusion and the attention-pooling weights agree on a low band around the center of the clip for adventitious events, whereas integrated gradients concentrates on single patches. For normal events, occlusion attributes most of the effect to bands above the filter cut-off.

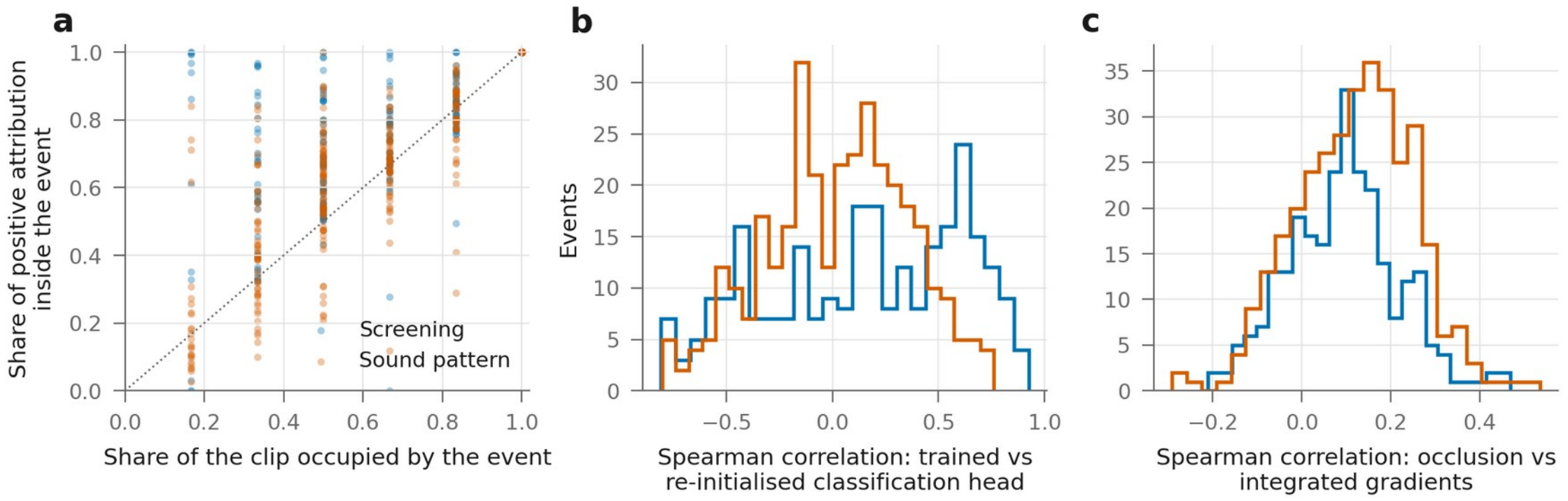

**Figure S6.** Per-event faithfulness statistics for correctly classified events. a Share of positive occlusion attribution inside the annotated event against the share of the clip that the event occupies (dotted line: equality). b Correlation between occlusion maps of the trained model and of a model whose classification head was re-initialized. c Correlation between occlusion and integrated-gradients maps.

## Table S16 Summary of faithfulness checks (correctly classified events)

| Measure | Screening | Sound pattern |
|---|---|---|
| Area under deletion curve, attribution-ordered | 0.605 | 0.507 |
| Area under deletion curve, random order | 0.719 | 0.688 |
| Difference (random minus attribution-ordered) | 0.114 | 0.182 |
| Share of attribution inside the annotated event | 0.80 | 0.67 |
| Share of the clip occupied by the event | 0.70 | 0.63 |
| Median correlation with re-initialized head | 0.19 | 0.03 |
| Median correlation, occlusion vs integrated gradients | 0.09 | 0.13 |

## S10 SPRSound releases and published event-classification results

The official SPRSound test set combines an intra-patient subset, drawn from the children of the training set, with an inter-patient subset of new children [2]; published event-level results on it (Table S17) are therefore partly record-wise. When the five best models of the 2022 challenge were re-scored on the blind 2023 test set, whose participant identifiers do not occur in the earlier releases, their Task 1-1 scores fell by 0.10 to 0.15 [24], a drop that may reflect new children as well as new annotators and recording conditions. Our screening outcome corresponds to Task 1-1; at the model's own decision (sensitivity 0.804, specificity 0.936) it reached a Score of 0.867 under patient-grouped cross-validation. The comparison is indicative only: our cohort pools all releases and excludes recordings without a documented diagnosis, the evaluation design differs, and our three-class sound-pattern outcome does not correspond to the seven classes of Task 1-2.

**Table S17 Published event-level results on SPRSound and their test data**

| Study | Test data | Children in test also in training | Task 1-1 Score | Task 1-2 Score |
|---|---|---|---|---|
| Zhang et al. 2022 [2], baseline | Official 2022 test | Partly | 0.752 | 0.616 |
| | Intra-patient subset (testing-1) | Yes | 0.722 | 0.589 |
| | Inter-patient subset (testing-2) | No | 0.774 | 0.636 |
| Li et al. 2022 [25] | Official 2022 test | Partly | 0.889 | 0.820 |
| Chen et al. 2022 [26] | Official 2022 test | Partly | 0.89 | 0.80 |
| Ngo et al. 2023 [27] | Official 2022 test | Partly | 0.849 | 0.774 |
| Five best 2022 models [24] | Blind 2023 test | No | 0.668–0.785 | 0.555–0.648 |
| Pessoa et al. 2023 [28] | Blind 2023 test | No | 0.756 | 0.467 |
| This study (screening) | Nested patient-grouped cross-validation | No | 0.867 | – |

Score: mean of the arithmetic and harmonic means of sensitivity and specificity, as defined for SPRSound [2]; Task 1-1, normal versus adventitious events; Task 1-2, seven event classes. Values as reported by the cited studies. "Children in test also in training" refers to the published split; for the blind 2023 test set it is judged from participant identifiers. The value for this study is not directly comparable (see text).

The analyzed cohort combines four release groups of the database, and no child appears in more than one (Table S18). The release groups differ sharply: adventitious events made up 24.1% of annotated events in the 2022 group and 5.3% in the 2024 detection test, and the median annotated duration was 1.2 s against 2.0 s. A cross-release validation (2022 to 2023, 86 test children; 2022 and 2023 to 2024, 322 test children) is feasible but was not performed: its result would mix new children with differences in annotation and recording, and, given the duration dependence of the operating point (Table S6), a lower sensitivity would be expected in the later releases.

### Table S18 Composition of the analyzed cohort by SPRSound release group

| Release group | Children (recordings; events) | Adventitious share of events | Annotated duration, median (IQR), ms | Pneumonia / bronchial / normal-other children |
|---|---|---|---|---|
| 2022: classification training set and 2022 test sets | 287 (2493; 9072) | 24.1% | 1221 (856–1620) | 192 / 58 / 37 |
| 2023: blind test | 86 (767; 2950) | 21.5% | 1587 (1176–2088) | 41 / 24 / 21 |
| 2024: detection test | 322 (1592; 7186) | 5.3% | 2048 (1631–2501) | 179 / 83 / 60 |
| 2025: BioCAS test | 41 (133; 485) | 19.8% | 2003 (1656–2451) | 26 / 14 / 1 |
| Total | 736 (4985; 19693) | 16.7% | – | 438 / 179 / 119 |

Analyzed cohort after exclusions. The 2022 group comprises the classification training set and the 2022 intra- and inter-patient test sets, whose intra-patient children are training children; no child appears in more than one group. Disease groups are counted per child; adventitious share is the proportion of annotated events with any adventitious sound.